\documentclass[%
 reprint,
 amsmath,amssymb,
 aps,
prb,
floatfix,
]{revtex4-1}

\usepackage{graphicx}
\usepackage{tikz-feynman,contour}
\tikzfeynmanset{compat=1.1.0}
\tikzfeynmanset{/tikzfeynman/momentum/arrow shorten = 0.3}
\tikzfeynmanset{/tikzfeynman/warn luatex = false}
\usepackage{subcaption}
\usepackage{dcolumn}
\usepackage{bm}
\usepackage{amssymb}
\usepackage{braket}
\usepackage{siunitx}
\usepackage{amsmath}
\usepackage{mathtools}
\usepackage{amsfonts}
\usepackage[breaklinks,colorlinks = true,linkcolor = magenta,urlcolor=magenta,citecolor=red]{hyperref}
\usepackage{xcolor}
\usepackage{dblfloatfix}

\newcommand{\ICTS}{International Centre for Theoretical Sciences, Tata Institute of Fundamental Research, Bangalore 560089, India}

\begin{document}

\title{Elastic signatures of a spin-nematic}

\author{Monica Bapna}
\email{monica.bapna@icts.res.in}
\affiliation{\ICTS}
\author{Subhro Bhattacharjee}
\email{subhro@icts.res.in}
\affiliation{\ICTS}

\date{\today}

\begin{abstract}
We study the elastic signatures-- renormalisation of sound velocity and magnetostriction -- of the spin-nematic  phase of a spin-$1$ magnet on a triangular lattice described by the bilinear-biquadratic spin Hamiltonian. We show that at low temperatures, the scattering of the acoustic phonons from the Goldstone modes of the nematic phase lead to a powerlaw renormalisation of the fractional change in the sound velocity, $v$, as a function of temperature, $T$, {\it i.e.} $\Delta v/v\propto T^3$ as opposed to the same in the high temperature paramagnet where $\Delta v/v\propto T^{-1}$. At the  generically discontinuous- nematic transition, there is a jump in magnetostriction as well as $\Delta v/v$ along with enhanced $h^4$ dependence on the magnetic field, $h$, near the nematic transition. These signatures can help positively characterise the spin-nematic in general and in particular the realisation of such a phase in the candidate material NiGa$_2$S$_4$.   
\end{abstract}

\maketitle


\section{\label{sec:Introduction}Introduction}

Interplay of symmetries and competing interactions can stabilise a plethora of  magnetic phases in spin systems with inconclusive experimental signatures for conventional probes. This not only include  issues of finding {\it smoking-gun} signatures of fractionalised quasi-particles in quantum spin liquids, but a much broader context pertaining to many other unconventional phases such as higher (than dipole) multi-pole orders in a variety of candidate systems with spin moments $S>1/2$.\cite{PhysRevLett.85.2188,chandra2002hidden,PhysRevB.86.184419,patri2019unveiling,harter2017parity,PhysRevB.101.155118} A possible resolution to the above conundrum  is to invoke a combination of  probes to gather complementary insights. An important class of such experimental probes like vibrational Raman and infrared scatterings as well as ultrasonic spectroscopy aim to exploit the ubiquitous magnetoelastic coupling to reveal the properties of the unconventional magnetic ground states and low energy excitations via the phonons.\cite{PhysRevLett.93.177402,PhysRevB.83.184421,PhysRevB.93.144412,PhysRevB.106.054507}  

In this paper, we shall develop the theory of magnetoelastic coupling for the rather elusive spin-nematic phases~\cite{blume1969biquadratic,PhysRevLett.27.1383,matveev1973quantum,PhysRevLett.66.100} in spin-1 magnets and apply it to predict its elastic signatures. In a spin-nematic, the ground state has no magnetic dipole moment but has a finite expectation value for the the quadrupole moment which is a bilinear of spins. Here, we shall consider only on-site spin-nematic characterised by the symmetric traceless operator\cite{PhysRevB.74.092406,penc2011spin} :
\begin{equation}
Q_i^{\mu\nu}=\frac{(S_i^{\mu}S_i^{\nu}+S_i^{\nu}S_i^{\mu})}{2}-\frac{S(S+1)\delta^{\mu\nu}}{3}, ~
\label{QuadrupoleDefinition}
\end{equation}
where $S^\alpha_i$ ($\alpha=x,y,z$) are spin-1 operators at lattice site $i$. Such order has been proposed for the triangular lattice magnet NiGa$_2$S$_4$~\cite{Nakatsuji1697,doi:10.1143/JPSJ.75.083701,PhysRevB.74.092406,PhysRevLett.97.087205,PhysRevB.79.214436} where it is stabilized by a sizeable biquadratic term $\sim ({\bf S}_i\cdot{\bf S}_j)^2$ that can arise from spin-lattice coupling.\cite{PhysRev.120.335,PhysRevB.12.2710,PhysRevB.74.092406}

${\text{NiGa}_2\text{S}_4}$ is a layered material where $\text{Ni}^{2+}$ forms an isotropic  triangular lattice with $S=1$ at each site. The system fails to show any conventional long range magnetic order in neutron scattering experiments to the lowest temperature measured (Curie Weiss temperature, $\theta_{W}=-80K$).\cite{Nakatsuji1697,nakatsuji2007coherent} However the state below $T\sim 10$ K shows $\sim T^2$ magnetic specific heat and constant magnetic susceptibility indicating the presence of low energy linearly dispersing excitations. It was  subsequently proposed that this spin-1 triangular lattice magnet could possibly have spin ferronematic~ \cite{PhysRevB.74.092406,PhysRevLett.97.087205,PhysRevB.79.214436} or  three sublattice~ nematic\cite{doi:10.1143/JPSJ.75.083701} ordering (see Fig. \ref{diagram_ugmodes}). This spin-nematic state forms the right starting point~\cite{PhysRevB.79.214436} to understand the relevance of the third nearest neighbour Heisenberg exchange~\cite{pradines2018study,PhysRevLett.105.037402,PhysRevB.76.140406,PhysRevLett.99.037203} as well as spin freezing below $10$ K~\cite{PhysRevLett.115.127202}-- both relevant for the material. However, the most pertinent experiments for our work are the recent Raman measurements probing the phonons in NiGa$_2$S$_4$~\cite{PhysRevLett.125.197201} which shows substantial spin-lattice coupling. We build on the above experimental indication of sizeable magnetoelastic  coupling to show that the possible spin-nematic can have a strong signature in the elastic sector -- namely the strain and the sound velocity -- which can form further experimental probes to such unconventional spin-quadrupolar order. These lattice signatures can be measured very accurately and have recently proved very useful in obtaining information about unconventional magnetic phases and phase transitions.\cite{PhysRevB.83.184421,tang2022spin}

Being bilinear in spins, the spin-nematic is time-reversal (TR) invariant, but, breaks the spin-rotation symmetry and has been dubbed as {\it moment-free magnetism}.\cite{PhysRevLett.66.100} Unlike dipolar ordering, such  quadrupolar ordering is hard to detect via  neutron scattering.\cite{PhysRevLett.70.2479}  However, the same TR even order parameter is expected to couple strongly to the lattice vibrations and hence provides a possible way to probe it. Further, even in absence of static dipole moment, the breaking of spin-rotation symmetry lead to gapless Goldstone modes -- spin-nematic waves. The coupling of such gapless modes to the acoustic phonons  further gives a way to detect the former in ultrasound experiments. 

Here, we report the effect of magnetoelastic coupling in a uniaxial spin-nematic by deriving the microscopics of such coupling and in particular the coupling of the acoustic phonons to the nematic Goldstone modes to obtain the renormalisation of sound speed at low temperatures deep inside the spin-nematic phase. We complement the microscopics approach with a phenomenological Landau-Ginzburg theory for the long wavelength dipolar, quadrupolar and strain modes to capture the effect of the thermal transition out of the spin-nematic phase on the magnetostriction and sound speed renormalisation. While we focus on  NiGa$_2$S$_4$~\cite{Nakatsuji1697,nakatsuji2007coherent}, our results are easily generalised to other cases of spin-nematic order.

The rest of this paper is organised as follows. In section \ref{sec:Heff}, we introduce the microscopic setting of magnetoelastic coupling (Eq. \ref{eq:spham}) in spin-1 magnets on triangular lattice with bilinear-biquadratic exchanges (Eq. \ref{Heff}) and summarize its different phases-- motivated by the low energy physics of NiGa$_2$S$_4$-- and discuss the effect of such coupling in fractional change of sound speed (Eq. \ref{dvbyv}). The above formalism is used to calculate the coupling between the nematic Goldstone modes and the acoustic phonons and hence the temperature dependence of fractional change in sound speed deep inside the  both the ferronematic and the three sublattice nematic phase in Section \ref{sec:dvbyvnematic}. Typically the temperature dependence of the sound speed is given by a powerlaw, $\Delta v/v\sim T^a$ where $a=3$ (Eq. \ref{FNdvbyv}) for a ferronematic and can vary between $1-3$ (Eq. \ref{eq_3sndvv3}) for the three sublattice nematic depending on the temperature range and the ratio of bilinear and biquadratic coupling $J/K$ (see Eq. \ref{Heff}). This is unlike the case of the thermal paramagnet where $\Delta v/v\propto 1/T$ (Eq. \ref{dvbyvpara}). Complementary to the microscopics, we study the phenomenological Landau-Ginzburg  theory for the long wavelength dipolar, quadrupolar and strain modes in Section \ref{sec:LandauTheory}. We find the symmetry allowed irreducible representations to obtain the mean field free energy which we use in Section \ref{Observables} to examine the elastic signatures -- fractional change in length (magnetostriction), $\Delta L/L$, and $\Delta v/v$ -- of the dipolar and quadrupolar ordering via symmetry allowed magnetoelastic couplings. We find that magnetic(nematic) ordering would lead to continuous change(jumps) in $\Delta L/L$ and $\Delta v/v$ along with an enhanced $h^4$ dependence on the magnetic field, $h$, near the nematic transitions. Observing jumps in fractional change in length and fractional change in sound speed in the absence of any magnetization would strongly favour the case for a spin-nematic. Various details are summarised in appendices.

\section{\label{sec:Heff}Spin-lattice coupling in Spin-1 magnet}

The starting point is the nearest neighbour minimal spin-1 bilinear-biquadratic model on the triangular lattice~\cite{PhysRevB.74.092406,penc2011spin} given by
\begin{equation}
H=J\sum_{\langle ij\rangle}(\mathbf{S}_{i}\cdot\mathbf{S}_{j})+K\sum_{\langle ij\rangle}(\mathbf{S}_{i}\cdot\mathbf{S}_{j})^2 
\label{Heff}
\end{equation}
where in addition to the usual (first) Heisenberg term, we also have the (second) biquadratic spin exchanges. These higher order exchanges can be obtained from an underlying higher energy multi-orbital Hubbard model.\cite{fazekas1999lecture} Notably, the biquadratic term can be further renormalised by magnetoelastic coupling by integrating out the phonons~\cite{PhysRevB.74.092406,PhysRevB.66.064403} whose importance is evident in the recent Raman scattering experiments.\cite{PhysRevLett.125.197201} Both virtual hopping and phonon effects naturally give rise to $K<0$ while  $J>0$ from the former.  

It is useful to re-write the above Hamiltonian, up to a constant, using the spin-$1$ operator identity~\cite{penc2011spin} : $(\mathbf{S}_i\cdot\mathbf{S}_j)^2=\frac{1}{3}S^2(S+1)^2-\frac{1}{2}(\mathbf{S}_i\cdot\mathbf{S}_j)+\sum_{\mu,\nu}Q_i^{\mu\nu}Q_j^{\mu\nu}$ as $H=\left(J-\frac{K}{2}\right)\sum_{\langle ij\rangle}\mathbf{S}_{i}\cdot\mathbf{S}_{j}+K\sum_{\langle ij\rangle;\mu,\nu}Q^{\mu\nu}_iQ^{\mu\nu}_j$. Eq. \ref{Heff} is clearly invariant under global spin rotations as well as various symmetries of the triangular lattice and TR. The ground states, however, spontaneously  break different symmetries depending on the ratio of $K/J$.

\begin{figure}
\centering
\includegraphics[width=\linewidth,keepaspectratio]{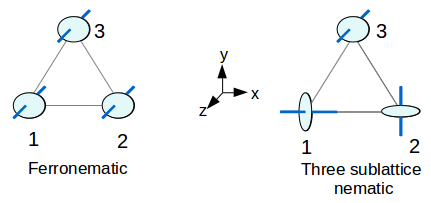}
\caption{Schematic : The uniaxial ferro (Eq. \ref{SN1}) and the three sublattice (Eq. \ref{Eq_3snop}) spin-nematic orders on the triangular lattice. The directors are shown in deep blue.}
\label{diagram_ugmodes}
\end{figure}

Most important to us is the large $K/J$ regime where an uniaxial uniform (ferro) spin-nematic is stabilised~\cite{PhysRevB.74.092406} for $K<0$. For spin-1, an on-site uniaxial nematic state is stabilized (this is clear from the discussion on the spin-1 wave functions in Refs. \onlinecite{PhysRevLett.97.087205, penc2011spin,articleTamasToth} as summarized in Appendix \ref{wavefunctions}) where the director of the nematic is uniform at all the lattice sites as shown in Fig. \ref{diagram_ugmodes}. The order parameter that characterise such a ferronematic order is 
\begin{equation}
\langle {Q}^{\mu\nu}\rangle=\mathcal{Q}_{FN}\bigg(n^{\mu}n^{\nu}-\frac{1}{3}\delta^{\mu\nu}\bigg) ~.
\label{SN1}
\end{equation}
where $\mathcal{Q}_{FN}$ and $\hat{\mathbf{n}}$ are respectively the  magnitude  and director of the ferronematic. Indeed, for $J>0$ and $K<0$, the ferronematic (spiral) phase is stable for $|K|/J>2(<2)$ within mean-field analysis.\cite{PhysRevB.79.214436}

While the microscopic mechanisms as discussed above, favours $K<0$, it is useful to consider the case of $K>0$, whence for  large $K/J$ a  three sublattice nematic is stabilised in the triangular lattice where the directors of the spin-nematic are orthogonal to each other on the three sublattices~\cite{doi:10.1143/JPSJ.75.083701} as shown in Fig. \ref{diagram_ugmodes}. The relevant mean-field wave-functions for the three sublattice nematic orders are briefly summarised in Appendix \ref{wavefunctions}. 

Turning the the limit of $J/|K|\gg 1$, on a triangular lattice, for $J>0$, a natural competing (with the above two nematic) phase is the 120$^\circ$ coplanar spiral with a non-zero spin expectation given by
\begin{eqnarray}
\langle\mathbf{S}_i\rangle&&=m\hat{\mathbf{s}}_{i} ~~{\rm with}~~\hat{\mathbf{s}}_{i}=\text{cos}(\mathbf{q}\cdot\mathbf{r}_{i})\hat{\mathbf{e}}_x+\text{sin}(\mathbf{q}\cdot\mathbf{r}_{i})\hat{\mathbf{e}}_y
\label{spiral1}
\end{eqnarray}
where $m$ is the magnitude of magnetization, $\mathbf{q}=\frac{2\pi}{3}\hat{\mathbf{x}}+\frac{2\pi}{\sqrt{3}}\hat{\mathbf{y}}$ is the spiral wave vector and $\hat{\mathbf{e}}_x$ and $\hat{\mathbf{e}}_y$ are orthogonal unit vectors in the plane of spin ordering. Note that the spin ordering induces a parasitic quadrupole moment, $\mathcal{Q}_{par}\sim m^2$,  which should be distinguished from the pure spin-nematic ordering discussed above.   

\subsection{\label{sec:Coupling}\label{sec:dvbyv}Spin-phonon coupling}

The Raman scattering experiments~\cite{PhysRevLett.125.197201} indicate the presence of substantial magnetoelastic coupling in NiGa$_2$S$_4$,  possibly arising from the modulation of the spin exchange coupling constants ($J$ and $K$) in the Hamiltonian in Eq. \ref{Heff} by the phonons. This is obtained via Taylor expansion of the exchange constants in lattice displacements about their equilibrium positions as \cite{PhysRevB.83.184421}
\begin{eqnarray}
J_{ij}&=& J_{0}+\frac{\partial J_{ij}}{\partial \mathbf{R}_{ij}}\cdot\mathbf{R}_{ij}+\frac{1}{2}\mathbf{R}_{ij}\cdot\frac{\partial^2 J_{ij}}{\partial \mathbf{R}_{ij}^2}\cdot\mathbf{R}_{ij} \nonumber\\
K_{ij}&=& K_{0}+\frac{\partial K_{ij}}{\partial \mathbf{R}_{ij}}\cdot\mathbf{R}_{ij}+\frac{1}{2}\mathbf{R}_{ij}\cdot\frac{\partial^2 K_{ij}}{\partial \mathbf{R}_{ij}^2}\cdot\mathbf{R}_{ij} ~,
\label{eq_exchange}
\end{eqnarray}
where $\mathbf{R}_{ij}=\mathbf{R}_{i}-\mathbf{R}_{j}$ and $\mathbf{R}_{i}$ is the displacement of site $i$ from its equilibrium position $\mathbf{R}_{i}^{0}$.  Eq. \ref{Heff} then becomes
\begin{align}
    H &= H_{sp}+H_{sp-ph}
\label{HTotSpinPhonon}
\end{align}
where 
$H_{sp}$ is the spin Hamiltonian in Eq. \ref{Heff} with the exchanges now being given by the equilibrium values (first term in Eq. \ref{eq_exchange}) and $H_{sp-ph}$ being the spin-phonon coupling Hamiltonian of the form,
\begin{eqnarray}
 H_{sp-ph} &=& H_{1}+H_{2},
 \label{eq:spham}
 \end{eqnarray}
where
\begin{eqnarray}
 H_{1} &=& \sum_{\mathbf{k}} U_{\mathbf{k}}^{(1)} A_{\mathbf{k}}+\sum_{\mathbf{k}} \widetilde{U}_{\mathbf{k}}^{(1)} A_{\mathbf{k}} \label{HspinphononH1}\\
 H_{2} &=& \frac{1}{2}\sum_{\mathbf{k},\mathbf{k}^{\prime}} U_{\mathbf{k},\mathbf{k}^{\prime}}^{(2)} A_{\mathbf{k}} A_{-\mathbf{k}^{\prime}}+\frac{1}{2}\sum_{\mathbf{k},\mathbf{k}^{\prime}} \widetilde{U}_{\mathbf{k},\mathbf{k}^{\prime}}^{(2)} A_{\mathbf{k}} A_{-\mathbf{k}^{\prime}}
\label{HspinphononH2}
\end{eqnarray}
are respectively linear and quadratic in lattice displacement operator given by $A_{\mathbf{k}} = a_{\mathbf{k}} + a_{-\mathbf{k}}^{\dagger}$  with $a^{\dagger}_{\mathbf{k}}$ being the bosonic phonon creation operator. $U_{\mathbf{k}}^{(1)}$, $\widetilde{U}_{\mathbf{k}}^{(1)}$, 
 $U_{\mathbf{k},\mathbf{k}^{\prime}}^{(2)}$ and $\widetilde{U}_{\mathbf{k},\mathbf{k}^{\prime}}^{(2)}$ depend on the spin operators and are given by
\begin{eqnarray}
&& U_{\mathbf{k}}^{(1)} = \sum_{\langle ij\rangle} (\mathbf{S}_{i}\cdot\mathbf{S}_{j}) \frac{(e^{i\mathbf{k}\cdot{\mathbf{R}_{i}^{0}}}-e^{i\mathbf{k}\cdot{\mathbf{R}_{j}^{0}}})}{\sqrt{2 M N \omega_{0,\mathbf{k}}^{ph}}}\left(\frac{\partial J_{ij}}{\partial \mathbf{R}_{ij}}\cdot\mathbf{e}_{\mathbf{k}}\right) \nonumber\\
&& \widetilde{U}_{\mathbf{k}}^{(1)} = \sum_{\langle ij\rangle} (\mathbf{S}_{i}\cdot\mathbf{S}_{j})^2 \frac{(e^{i\mathbf{k}\cdot{\mathbf{R}_{i}^{0}}}-e^{i\mathbf{k}\cdot{\mathbf{R}_{j}^{0}}})}{\sqrt{2 M N \omega_{0,\mathbf{k}}^{ph}}}\left(\frac{\partial K_{ij}}{\partial \mathbf{R}_{ij}}\cdot\mathbf{e}_{\mathbf{k}}\right)\nonumber\\
\label{U1}
\end{eqnarray}
and
\begin{align}
& U_{\mathbf{k},\mathbf{k}^{\prime}}^{(2)} = \sum_{\langle ij\rangle}\frac{(\mathbf{S}_{i}\cdot\mathbf{S}_{j})}{2 M N \sqrt{\omega_{0,\mathbf{k}}^{ph} \omega_{0,-\mathbf{k}^{\prime}}^{ph}}} (e^{i\mathbf{k}\cdot{\mathbf{R}_{i}^{0}}}-e^{i\mathbf{k}\cdot{\mathbf{R}_{j}^{0}}})\nonumber\\
&~~~~~~~~~~~~~~~\times \left(\mathbf{e}_{-\mathbf{k}^{\prime}}\cdot\frac{\partial^2 J_{ij}}{\partial{\mathbf{R}_{ij}^{2}}}\cdot\mathbf{e}_{\mathbf{k}}\right)(e^{-i\mathbf{k}^{\prime}\cdot{\mathbf{R}_{i}^{0}}}-e^{-i\mathbf{k}^{\prime}\cdot{\mathbf{R}_{j}^{0}}}) \nonumber\\
& \widetilde{U}_{\mathbf{k},\mathbf{k}^{\prime}}^{(2)} = \sum_{\langle ij\rangle} \frac{(\mathbf{S}_{i}\cdot\mathbf{S}_{j})^2}{2 M N \sqrt{\omega_{0,\mathbf{k}}^{ph} \omega_{0,-\mathbf{k}^{\prime}}^{ph}}} (e^{i\mathbf{k}\cdot{\mathbf{R}_{i}^{0}}}-e^{i\mathbf{k}\cdot{\mathbf{R}_{j}^{0}}})\nonumber\\
&~~~~~~~~~~~~~~~\times \left(\mathbf{e}_{-\mathbf{k}^{\prime}}\cdot\frac{\partial^2 K_{ij}}{\partial{\mathbf{R}_{ij}^{2}}}\cdot\mathbf{e}_{\mathbf{k}}\right)(e^{-i\mathbf{k}^{\prime}\cdot{\mathbf{R}_{i}^{0}}}-e^{-i\mathbf{k}^{\prime}\cdot{\mathbf{R}_{j}^{0}}})
\label{U2}
\end{align}
where $M$ and $N$ are the mass and number of a nickel ions respectively, $\mathbf{e}_{\mathbf{k}}$ is the phonon polarization vector, $\omega_{0,\mathbf{k}}^{ph}$ is the bare phonon frequency corresponding to the Harmonic phonon Hamiltonian,
\begin{align}
    H_{ph} =& \sum_{\mathbf{k}} \omega_{0,\mathbf{k}}^{ph} a^{\dagger}_{\mathbf{k}} a_{\mathbf{k}}
    \label{eq_phonon_harmonic}
\end{align}
For acoustic phonons at long wavelengths (${\bf k}\rightarrow 0$) we have $\omega_{0,\mathbf{k}}^{ph} \approx v_0 |\mathbf{k}|$ where $v_0$ is the bare sound speed. Above and in the rest of this work, we have set $\hbar=1$.

\subsection{Fractional Change in the sound speed}

The effect of the spin-phonon coupling on the fractional change in sound speed is obtained from the real part of the phonon self-energy, $\Sigma (\mathbf{q},\omega)$,  as,\cite{PhysRevB.83.184421}
\begin{equation}
\frac{\Delta v}{v} =  \lim_{\mathbf{q} \xrightarrow{} 0} \frac{\Delta \omega_{\mathbf{q}}^{ph}}{\omega_{0,\mathbf{q}}^{ph}} \approx \lim_{\mathbf{q} \xrightarrow{} 0} \frac{Re \Sigma (\mathbf{q},\omega_{0,\mathbf{q}}^{ph})}{\omega_{0,\mathbf{q}}^{ph}}.
\label{dvbyv}
\end{equation}

The phonon self-energy due to the interactions with the spins, in turn, can be obtained from the phonon propagator $G^{ph} (\mathbf{q},\tau-\tau^{\prime})=-\langle T_{\tau}(A_{\mathbf{q}}(\tau)A_{-\mathbf{q}}(\tau^{\prime}))\rangle$ in Matsubara space and is given by the Dyson equation
\begin{align}
    G^{ph} (\mathbf{q},i \Omega_{n})=\frac{1}{\left(G^{ph}_{0} (\mathbf{q},i \Omega_{n})\right)^{-1}- \Sigma (\mathbf{q},i \Omega_{n}) }
    \label{eq_dressedgnf}
\end{align}
where $\Omega_{n}$ are the bosonic Matsubara frequencies and the bare phonon propagator given by
\begin{align}
G^{ph}_{0} (\mathbf{q},i \Omega_{n})=\frac{2 \omega_{0,\mathbf{q}}^{ph} }{(i \Omega_{n})^2-(\omega_{0,\mathbf{q}}^{ph})^2}.
\end{align}

The phonon self energy due to the magnetoelastic coupling would depend on correlations of the spins and hence fractional change in sound speed can be used to probe the spin physics. In the following sections we use this formalism to calculate the temperature dependence of $\Delta v/v$ in the ferronematic, three sublattice nematic and the thermal paramagnet.

\section{\label{sec:dvbyvnematic}Fractional change in sound speed in spin-nematic phase}

Deep inside the spin-nematic phase -- both ferro and three sublattice, the renormalisation of the sound speed is brought about by the interaction between the acoustic phonons and the spin-nematic waves~\cite{PhysRevB.10.4650} via the coupling given by Eq. \ref{eq:spham}. To this end we use the {\it spin-nematic wave} theory for the ferronematic (Matveev {\it et. al.}\cite{matveev1973quantum}) and three sublattice nematic (Tsunetsugu {\it et. al.}\cite{doi:10.1143/JPSJ.75.083701}) to write the spins in terms of the low energy Goldstone bosons of the respective nematic orders -- summarised in Appendix \ref{sec:AppendixSW} for completeness. 

\subsection{\label{sec:dvbyvFerronematic}The Ferronematic phase}
To get the temperature dependence of $\Delta v/v$ (Eq. \ref{dvbyv}) in the ferronematic, we calculate phonon self energy starting with the following Hamiltonian for the phonon and linear spin-nematic waves
\begin{align}
    \mathcal{H}_{FN} &= H_{ph}+H_{sp,FN}+H_{sp-ph,FN} ~,
\label{HTotFN}
\end{align}
obtained from Eqs. \ref{HTotSpinPhonon} and  \ref{eq_phonon_harmonic}. Here $H_{ph}$ is the harmonic phonon Hamiltonian (Eq. \ref{eq_phonon_harmonic}), $H_{sp,FN}$ and $H_{sp-ph,FN}$ are respectively the linear spin-nematic wave Hamiltonian for the ferronematic phase and  the phonon-spin-nematic wave coupling Hamiltonian respectively whose form we now discuss.

$H_{sp,FN}$ can be obtained from Eq. \ref{Heff} by expressing the spins in terms of boson operators that create spin-nematic wave.\cite{matveev1973quantum} This is done for the ferronematic state by introducing two bosons at every site $i$, with creation operators given by $b_{i1}^{\dagger}$ and $b_{i2}^{\dagger}$ that capture deviations from the mean field ferronematic ground state. For a mean field state with the director along the $\hat{\bf z}$-direction, the wave function is $\otimes_{i}\ket{0}_{i}$  (Appendix \ref{wavefunctions}) and $b_{i1}^{\dagger}\ket{0}_{i}=\ket{1}_{i},b_{i1}\ket{1}_{i}=\ket{0}_{i},b_{i2}^{\dagger}\ket{0}_{i}=\ket{\Bar{1}}_{i}$ and $b_{i2}\ket{\Bar{1}}_{i}=\ket{0}_{i}$. The relation between the spin operators and the above bosons are given in Appendix \ref{sec:AppendixSWFN}. Expressing the spin operators in terms of the bosons in Eq. \ref{Heff} and approximating to the harmonic order we get 
\begin{align}
    H_{sp,FN} &= \sum_{\mathbf{k}} \omega^s_{\mathbf{k}} \psi_{\mathbf{k}}^{\dagger} \psi_{\mathbf{k}} ~, 
\label{HspFN}
\end{align}
 with $\psi_{\mathbf{k}} = (d_{\mathbf{k},1},d^{\dagger}_{-\mathbf{k},2})$ where $d_{\mathbf{k},1}$ and $d_{\mathbf{k},2}$ are related to $(b_{\mathbf{k},1},b_{\mathbf{k},2})$ via Bogoliubov transformation (Eq. \ref{BogoDefn}). The dispersion is given by
 \begin{align}
    \omega^s_{\mathbf{k}} &= 6 K_0 \sqrt{(1-\gamma_{\mathbf{k}})\left(1+\gamma_{\mathbf{k}}-\frac{2J_0}{K_0} \gamma_{\mathbf{k}}\right)} ~,
\label{HspFNdispersion}
\end{align}
where $\gamma_{\mathbf{k}}=\frac{1}{6}\sum_{\boldsymbol{\delta}} e^{i \mathbf{k}\cdot\boldsymbol{\delta}}$ and  $\boldsymbol{\delta}$ is the distance to the six nearest neighbours. In the long wavelength limit, $\omega^s_{\mathbf{k}} \approx \Bar{c}_s |\mathbf{k}|$ with $\Bar{c}_s$ being the  ferro spin-nematic wave speed.

\begin{figure}
  \begin{subfigure}[t]{0.49\columnwidth}
    \begin{tikzpicture}
    \begin{feynman}
    \vertex (a);
    \vertex [right=1cm of a] (b);
    \vertex [above right=1.25cm of b] (f1);
    \vertex [below right=1.25cm of b] (f2);
    \diagram* {
    (a) -- [plain, black,  edge label'=\(\mathbf{k}\)] (b) -- [charged scalar, black, edge label=\(\mathbf{{k+q_2}}\)] (f1),
    (f2)-- [charged scalar, black, edge label'=\(\mathbf{{q_2}}\)] (b)
    };
    \end{feynman}
    \end{tikzpicture}
  \end{subfigure}
  \begin{subfigure}[t]{0.49\columnwidth}
    \begin{tikzpicture}
    \begin{feynman}
    \vertex (a);
    \vertex [above right=1.25cm of a] (f1);
    \vertex [below right=1.25cm of a] (f2);
    \vertex [above left=1.25cm of a] (f3);
    \vertex [below left=1.25cm of a] (f4);
    \diagram* {
    (a) -- [plain, black,  edge label'=\(\mathbf{k}\)] (f3),
    (a) -- [plain, black,  edge label'=\(\mathbf{-k'}\)] (f4),
    (a)-- [charged scalar, black, edge label'=\(\mathbf{{k-k'+q_2}}\)] (f1),
    (f2)-- [charged scalar, black, edge label'=\(\mathbf{{q_2}}\)] (a)
    };
    \end{feynman}
    \end{tikzpicture}
 \end{subfigure}
\caption{Interactions between the phonon and ferronematic spin waves (Eq. \ref{H12}). The left diagram comes from $H_{1,FN}$ and right diagram from first term in $H_{2,FN}$. Solid lines represent phonons and dashed lines represent ferronematic spin waves ($\psi$ in Eq. \ref{HspFN})}.
\label{vertex} 
\end{figure}
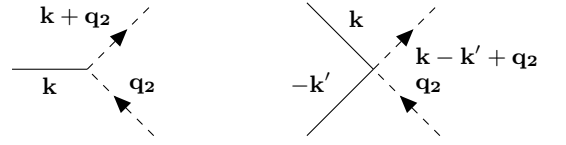

The coupling between the phonons and the ferro spin-nematic waves is obtained by using the same bosonic representation of the spin operators in  Eq. \ref{eq:spham}. The resultant Hamiltonian is given by
\begin{align}
   H_{sp-ph,FN} =H_{1,FN} + H_{2,FN} 
   \label{eq_fnsph}
\end{align}
where 
\begin{align}
    H_{1,FN} =& \sum_{\mathbf{k},\mathbf{q_2}} \psi^{\dagger}_{\mathbf{k}+\mathbf{q_2}} \mathcal{M}_{\mathbf{k},\mathbf{q_2}}^{(1)} \psi_{\mathbf{q_2}} A_{\mathbf{k}}\nonumber\\
H_{2,FN} =& \sum_{\mathbf{k},\mathbf{k}^{\prime},\mathbf{q_2}} \psi^{\dagger}_{\mathbf{k}-\mathbf{k}^{\prime}+\mathbf{q_2}} \mathcal{M}_{\mathbf{k},\mathbf{k}^{\prime},\mathbf{q_2}}^{(2)} \psi_{\mathbf{q_2}} A_{\mathbf{k}} A_{-\mathbf{k}^{\prime}} \nonumber\\
&+\sum_{\mathbf{k}} \mathcal{M}_{\mathbf{k}}^{(2),0} A_{\mathbf{k}} A_{-\mathbf{k}}
\label{H12}
\end{align}
represent respectively the linear (Eq. \ref{HspinphononH1}) and quadratic (Eq. \ref{HspinphononH2}) coupling with the phonons as shown in Fig. \ref{vertex}. The detailed expression of the scattering vertices $\mathcal{M}_{\mathbf{k},\mathbf{q_2}}^{(1)}$, $\mathcal{M}_{\mathbf{k},\mathbf{k}^{\prime},\mathbf{q_2}}^{(2)}$ and $\mathcal{M}_{\mathbf{k}}^{(2),0}$ are given in Appendix \ref{sec:AppendixTinFN}. Notably the second term in $H_{2,FN}$, given by $\mathcal{M}^{(2),0}_{\bf k}$ only depends on the displacement operators via a quadratic form similar to that of the Harmonic potential. This corresponds to the renormalisation of the bare phonon frequency and hence to sound speed due to ferronematic ordering.

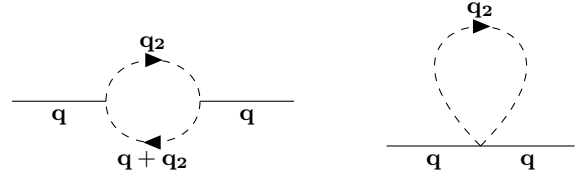
\begin{figure}
  \centering
  \begin{subfigure}[t]{0.49\columnwidth}
    \centering
    \begin{tikzpicture}
    \begin{feynman}
    \vertex (a);
    \vertex [right=1.25cm of a] (b);
    \vertex [right=1.25cm of b] (c);
    \vertex [right=1.25cm of c] (d);
    \diagram* {
    (a) -- [plain, black,  edge label'=\(\mathbf{q}\)] (b) -- [charged scalar, black,half left,looseness=1.5, edge label=\(\mathbf{{q_2}}\)] (c)-- [charged scalar, black,half left,looseness=1.5, edge label=\(\mathbf{{q+q_2}}\)] (b),
    (c)-- [plain, black,  edge label'=\(\mathbf{q}\)] (d)
    };
    \end{feynman}
    \end{tikzpicture}
  \end{subfigure}
  \begin{subfigure}[t]{0.49\columnwidth}
    \centering
    \begin{tikzpicture}
    \begin{feynman}
    \vertex (a);
    \vertex [right=1.25cm of a] (b);
    \vertex [right=1.25cm of b] (c);
    \diagram* {
    (a) -- [plain, black,  edge label'=\(\mathbf{q}\)] (b) --  [plain, black,  edge label'=\(\mathbf{q}\)] (c),
    b -- [charged scalar,black, out=135, in=45, loop, min distance=3cm,edge label=\(\mathbf{{q_2}}\)] b,
    };
    \end{feynman}
    \end{tikzpicture}  
 \end{subfigure}
\caption{Diagrams used to calculate contribution to the phonon self energy due to nematic fluctuations (Fig. \ref{vertex}). The left diagram comes from vertices due to $H_{1,FN}$ and right diagram from $H_{2,FN}$ (Eq. \ref{H12}).}.
\label{diagrams} 
\end{figure}

The contributions from the spin-nematic waves of the ferronematic to $\Delta v/v$ can be captured to the leading order in magnetoelastic coupling by calculating the phonon free energy due to $\mathcal{M}_{\mathbf{k},\mathbf{q_2}}^{(1)}$ and  $\mathcal{M}_{\mathbf{k},\mathbf{k}^{\prime},\mathbf{q_2}}^{(2)}$ terms given by Feynman diagrams in Fig. \ref{diagrams}. 

The phonon self energy (in Eq. \ref{eq_dressedgnf}) due to the above two contributions is then given by

\begin{align}
    \Sigma(\mathbf{q}, i\Omega_n)=\Sigma_1(\mathbf{q}, i\Omega_n)+\Sigma_2(\mathbf{q}, i\Omega_n)
\end{align}
where
\begin{eqnarray}
\Sigma_{1} (\mathbf{q},i \Omega_{n})&\approx&-\frac{1}{\beta}\sum_{\substack{\mathbf{q}_2,\lambda,\\\rho,\omega_1}}\mathcal{M}_{-\mathbf{q},\mathbf{q}+\mathbf{q}_2}^{(1),\lambda\rho} \mathcal{M}_{\mathbf{q},\mathbf{q}_2}^{(1),\rho\lambda}G^{m,\lambda\lambda}_{0}(\mathbf{q}_2,i \omega_1)\nonumber\\
&&~~~~~~~~~~~\times G^{m,\rho\rho}_{0}(\mathbf{q}+\mathbf{q}_2,i \Omega_{n}+i \omega_1) \nonumber\\
\Sigma_{2} (\mathbf{q},i \Omega_{n}) &\approx& 2 \mathcal{M}_{\mathbf{q}}^{(2),0}+\sum_{\mathbf{q}_2,\lambda} 2 \mathcal{M}_{\mathbf{q},\mathbf{q},\mathbf{q_2}}^{(2),\lambda\lambda} \langle \psi^{\dagger \lambda}_{\mathbf{q}_2}\psi^{\lambda}_{\mathbf{q}_2}\rangle~,
\label{MatsubaraSelfEnergyFN}
\end{eqnarray}
with $\lambda,\rho=1,2$ and $G^{m,\lambda\rho}_{0}$ denotes the bare Matrix ferronematic spin-wave Green's function defined as $G^{m,\lambda\rho}_{0} (\mathbf{q},\tau-\tau^{\prime})=-\langle T_{\tau}(\psi^{\lambda}_{\mathbf{q}}(\tau)\psi^{\dagger \rho}_{\mathbf{q}}(\tau^{\prime}))\rangle$. 
Due to the diagonal form of the spin wave Hamiltonian (Eq. \ref{HspFN}), the matrix is given by
\begin{equation}
G^{m}_{0} (\mathbf{q},i \Omega_{n})=\left(\begin{matrix} \frac{1}{i \Omega_{n}-\omega^s_{\mathbf{q}}} & 0 \\ 0 & \frac{-1}{i \Omega_{n}+\omega^s_{-\mathbf{q}}}\end{matrix}\right)    ~.
\end{equation}

The sum over the bosonic Matsubara frequencies in Eq. \ref{MatsubaraSelfEnergyFN} is evaluated using the standard techniques \cite{bruus2004many} and using Eq. \ref{dvbyv}, the temperature dependence of the fractional change in sound speed is obtained as
\begin{equation}
\frac{\Delta v}{v}=\Bar{c}_1+\Bar{c}_2 T^3 ~,
\label{FNdvbyv}
\end{equation}
The factors $\Bar{c}_1$ and $\Bar{c}_2$ (see expressions in Appendix \ref{sec:AppendixTinFN})  have contributions from first and second derivatives of both the bilinear and the biquadratic couplings and depend on the details of the lattice via the phonon spectrum and polarisation. However, the above temperature dependence is generically valid for other two dimensional lattices. The sound attenuation, $\propto Im\Sigma/v_0$, can similarly be calculated from the imaginary part of the phonon self energy. At this order only $\Sigma_1$ contributes and such sound attenuation generically proportional, at low frequencies, the to the phonon frequency.\cite{cottam1974spin,PhysRevResearch.1.033065} 

\subsection{\label{sec:dvbyv3SN}The three sublattice nematic phase}

The $\Delta v/v$ for the three sublattice nematic can be obtained in a similar way. In analogy with the ferronematic case (Eq. \ref{HTotFN}), the relevant Hamiltonian is given by
\begin{align}
    \mathcal{H}_{3SN} &= H_{ph}+H_{sp,3SN}+H_{sp-ph,3SN} ~, 
\label{HTot3SN}
\end{align}
where $H_{ph}$ is the harmonic phonon Hamiltonian (Eq. \ref{eq_phonon_harmonic}), $H_{sp,3SN}$ and $H_{sp-ph,3SN}$ are respectively the linear spin-nematic wave  Hamiltonian for the three sublattice nematic phase and  the phonon-three sublattice spin-nematic wave coupling Hamiltonian. 

Due to the non-uniform structure of this nematic phase, the calculations are somewhat more tedious and the relevant parts are relegated to Appendix \ref{sec:AppendixTin3SN}. The difference, however, in this case stems from the three sublattice spin-nematic wave spectrum. The spin-nematic wave theory about a three sublattice nematic state,  \cite{doi:10.1143/JPSJ.75.083701} such as $\prod_{\mathbf{\widetilde{R}}} \ket{x}_{\mathbf{\widetilde{R}},1} \ket{y}_{\mathbf{\widetilde{R}},2} \ket{z}_{\mathbf{\widetilde{R}},3}$ 
is obtained as follows. Here $\mathbf{\widetilde{R}}$ stands for the three sublattice unit cell and 1, 2 and 3 denote the sublattices of each unit cell with their nematic directors being along the $\hat{\mathbf{x}}$, $\hat{\mathbf{y}}$ and $\hat{\mathbf{z}}$ directions respectively as shown in Fig. \ref{diagram_ugmodes}. We introduce two bosons $\widetilde{\alpha}^\dagger$ and $\widetilde{\beta}^\dagger$ at each sublattice to capture the deviations from the ground state, {\it e.g.}, for sublattice-3, $\ket{z}_{\mathbf{\widetilde{R}},3} \equiv \ket{\text{vac}}_{\mathbf{\widetilde{R}},3}$ and $\ket{S_z=\pm 1}_{\mathbf{\widetilde{R}},3} =\frac{1}{\sqrt{2}}(\widetilde{\alpha}^\dagger_{\mathbf{\widetilde{R}},3} \pm i \widetilde{\beta}^\dagger_{\mathbf{\widetilde{R}},3})\ket{\text{vac}}_{\mathbf{\widetilde{R}},3}$. The details are summarised in Appendix \ref{sec:AppendixSW3SN}.

The resultant diagonalised harmonic Hamiltonian (similar to Eq. \ref{HspFN} for the ferronematic) is given by 

\begin{equation}
    H_{sp,3SN} = \sum_{\mathbf{k}, \widetilde{\lambda} \widetilde{\rho}} \bigg[\widetilde{\omega}^s_{+,\mathbf{k}}\text{\ }\widetilde{d}_{+, \mathbf{k}, \widetilde{\lambda} \widetilde{\rho}}^{~\dagger} \widetilde{d}_{+, \mathbf{k}, \widetilde{\lambda} \widetilde{\rho}} + \widetilde{\omega}^s_{-,\mathbf{k}} \text{\ }\widetilde{d}_{-, \mathbf{k}, \widetilde{\lambda} \widetilde{\rho}}^{~\dagger} \widetilde{d}_{-, \mathbf{k}, \widetilde{\lambda} \widetilde{\rho}} \bigg]
\label{Hsp3SN}
\end{equation}
where $\widetilde{\lambda} \widetilde{\rho} =\{12,23,31\}$, $\widetilde{d}_{\pm, \mathbf{k}, \widetilde{\lambda}\widetilde{\rho}}$ are the diagonalised bosonic annihilation operators with dispersions 
\begin{align}
\widetilde{\omega}^s_{\pm,\mathbf{k}}= 3 K_0\sqrt{(1 \pm |\widetilde{\gamma}_{\mathbf{k}}|)\left(1 \pm \bigg(1-\frac{2J_0}{K_0}\bigg) |\widetilde{\gamma}_{\mathbf{k}}|\right)} ~,
\label{dispersion3SN}
\end{align}
where $\widetilde{\gamma}_{\mathbf{k}} = \frac{1}{3} \sum_{\boldsymbol{\widetilde{\delta}}} e^{i \mathbf{k}.\boldsymbol{\widetilde{\delta}}}$ and $\boldsymbol{\widetilde{\delta}}=\{-\hat{\mathbf{x}}/2-\sqrt{3}\hat{\mathbf{y}}/2,\hat{\mathbf{x}},-\hat{\mathbf{x}}/2+\sqrt{3}\hat{\mathbf{y}}/2\}$ (lattice constant set to unity).

For $J_0 \neq K_0$, the $\widetilde{\omega}^s_{+,\mathbf{k}}$ branch is gapped. For $J_0/K_0\ll 1$ the $\widetilde{\omega}^s_{+,\mathbf{k}}$ branch lies entirely above the $\widetilde{\omega}^s_{-,\mathbf{k}}$ branch (touching $\widetilde{\omega}^s_{-,\mathbf{k}}$ at the Brillouin zone corners.\cite{doi:10.1143/JPSJ.75.083701}) Hence its effect can be neglected at low temperatures which is dominated by the long wavelength behaviour of the gapless mode $\widetilde{\omega}^s_{-,\mathbf{k}}$. The latter shows, in turn, a crossover depending on the ratio of $J_0/K_0$. Eq. \ref{dispersion3SN}, for generic $J_0/K_0$, at long wavelengths,  $\widetilde{\omega}^s_{-,\mathbf{k}}\approx c_s|{\bf k}|$ (with $c_s = 3  \sqrt{\frac{J_0 K_0}{2}}$). However for $J_0=0$, the long wavelength scaling of the dispersion changes to $\widetilde{\omega}^s_{-,\mathbf{k}}\approx \widetilde{c}_s|{\bf k}|^2$ (with $\widetilde{c}_s=\frac{3K_0}{4}$ )\cite{doi:10.1143/JPSJ.75.083701} such that for  $J_0/K_0\ll 1$, there is a crossover scale $k^*=\sqrt{\frac{8J_0}{K_0}}$ (in inverse units of lattice length scale) below which the dispersion is approximately linear and above which it is approximate quadratic. This change of the dispersion as well as the crossover affects the temperature dependence of the $\Delta v/v$ in the three sublattice nematic depending on the ratio of $J_0/K_0$ (see below).

Finally, the three sublattice nematic spin wave-phonon Hamiltonian, $H_{sp-ph,3SN}$ (in eq. \ref{HTot3SN}) is obtained, similar to the ferronematic case, by writing the spin operators (Eq. \ref{spinSW3SN}) in terms of the $\widetilde{d}_{\pm, \mathbf{k}, \widetilde{\lambda} \widetilde{\rho}}$ bosons of Eq. \ref{Hsp3SN} and substituting in Eq. \ref{HspinphononH1}, \ref{HspinphononH2}, \ref{U1} and \ref{U2} (analogous to Eq. \ref{eq_fnsph}). This leads to Eq. \ref{eq_Hspph-3SN}. The fractional change of sound speed due to the three sublattice ordering is then obtained similar to the the ferronematic (details in Appendix \ref{sec:AppendixTin3SN}) and is given by
\begin{eqnarray}
    \frac{\Delta v}{v} &\approx& \lim_{\mathbf{q} \xrightarrow{} 0} \Bigg[ \int_0^{k^*}dk \int d\theta_{\mathbf{\hat{k}}} \frac{Re \widetilde{\Sigma}_{\mathbf{k}}\left( \widetilde{\omega}^s_{-,\mathbf{k}} \propto |\mathbf{k}|  \right)}{\omega_{0,\mathbf{q}}^{ph}} \nonumber\\
    &+& \int_{k^*}^\infty dk \int d\theta_{\mathbf{\hat{k}}} \frac{Re \widetilde{\Sigma}_{\mathbf{k}}\left( \widetilde{\omega}^s_{-,\mathbf{k}} \propto |\mathbf{k}|^2  \right)}{\omega_{0,\mathbf{q}}^{ph}} \Bigg] ~, 
    \label{eq_3sndvv2}
\end{eqnarray}
accounting for the crossover between the linear and quadratic dispersions at $k^*$. This approximately leads to, in the three sublattice nematic,  
\begin{align}
    \frac{\Delta v}{v}=\left\{\begin{array}{l}
    \widetilde{c}_1 + \widetilde{c}_2 T^3~~~~{\rm for}~\beta J_0\gg 1 \\
    \widetilde{c}_3 + \widetilde{c}_4 T^2~~~~{\rm for}~\beta J_0\sim 1\\
    \widetilde{c}_5 + \widetilde{c}_6 T~~~~~{\rm for}~\beta J_0 < 1\\
    \end{array}\right.
    \label{eq_3sndvv3}
\end{align}
where $\beta=1/(k_B T)$ and the details of the pre-factors are given in Appendix \ref{sec:AppendixTin3SN}. No such crossover behaviour is observed for the ferronematic case since the spin wave dispersion remains linear in $k$ even on setting $J_0=0$. 

\subsection{\label{sec:dvbyvParamagnetic} 
The thermal paramagnet}

In contrast to the low temperature nematic phases discussed above, for the high temperature paramagnetic phase, the spin dynamics is faster than the acoustic phonons and integrating out spins, gives the following effective interaction Hamiltonian for phonons,\cite{PhysRevB.83.184421}
\begin{equation}
H_{\text{eff}}=\sum_{\mathbf{k}} \omega_{0,\mathbf{k}}^{ph} a^{\dagger}_{\mathbf{k}} a_{\mathbf{k}}+\frac{1}{2}\sum_{\mathbf{k},\mathbf{k}^{\prime}} V_{\mathbf{k}\mathbf{k}^{\prime}}A_{\mathbf{k}}A_{-\mathbf{k}^{\prime}}
\end{equation}
where
\begin{equation}
V_{\mathbf{k}\mathbf{k}^{\prime}}=\langle \Bar{\Bar{U}}_{\mathbf{k}\mathbf{k}^{\prime}}^{(2)}\rangle-\beta \langle\langle\Bar{\Bar{U}}_{\mathbf{k}}^{(1)}\Bar{\Bar{U}}_{-\mathbf{k}^{\prime}}^{(1)}\rangle\rangle
\label{Veffpara}
\end{equation}
Here $\Bar{\Bar{U}}_{\mathbf{k}}^{(1)}=U_{\mathbf{k}}^{(1)}+\widetilde{U}_{\mathbf{k}}^{(1)}$ and $\Bar{\Bar{U}}_{\mathbf{k}\mathbf{k}^{\prime}}^{(2)}=U_{\mathbf{k}\mathbf{k}^{\prime}}^{(2)}+\widetilde{U}_{\mathbf{k}\mathbf{k}^{\prime}}^{(2)}$ (which have been defined in Eq. \ref{U1} and Eq. \ref{U2}) and $\langle\langle....\rangle\rangle$ denotes the connected correlators for the spins, averaged over a thermal ensemble. The self energy is given by :
\begin{equation}
\Sigma (\mathbf{q},i \Omega_{n}) = V_{\mathbf{q}\mathbf{q}}     
\end{equation}
So, calculating the leading order temperature dependence of $\Delta v/v$ would involve various spin correlators which are then calculated using high temperature series expansion. To the leading order in $1/T$, this gives
\begin{equation}
\frac{\Delta v}{v} =a_1 + \widetilde{a}_2 \beta ~,
\label{dvbyvpara}
\end{equation}
where, notably, a constant contribution arises from the biquadratic term such that 
\begin{equation}
    a_1 = \sum_{\boldsymbol{\delta}} \frac{1}{3 M v_0^2} (\hat{\mathbf{q}}.\boldsymbol{\delta})^2 \left(\mathbf{e}_{-\mathbf{q}}.\frac{\partial^2 K}{\partial{\boldsymbol{\delta}^{2}}}.\mathbf{e}_{\mathbf{q}}\right) ~,
\end{equation}
where $\boldsymbol{\delta}$ runs over the six nearest neighbours. It is evident that the constant arises only due to the biquadratic coupling and would be absent for the case of pure Heisenberg model.


\section{\label{sec:LandauTheory}The Landau-Ginzburg Theory}

The above microscopic approach works deep inside the nematic at low temperatures or in the thermal paramagnet at high temperatures. For general elastic responses for the entire phase diagram, a more phenomenological Landau -Ginzburg theory-- that accounts for the 120$^\circ$ spiral and the spin-nematic (both ferro and three sub-lattice) ordering as well as the elastic degrees of freedom-- is useful to account for the spin-lattice physics of the system.\cite{patri2019unveiling} Below, we construct this theory to derive the elastic signatures of a spin-nematic in context of the NiGa$_2$S$_4$. Our calculations are easily generalised to other situations, in particular the spin-orbit coupled multi-polar orders.\cite{voleti2022probing,patri2019unveiling}

We start with the symmetry analysis for the three fields -- the magnetic, the spin-nematic and elastic -- for the point group symmetry of ${\text{NiGa}_2\text{S}_4}$ which is $\text{D}_{3d}$. For our present calculations, we choose the largest unit cell for all the type of orderings discussed above-- a single triangle-- and systematically isolate the relevant symmetry allowed terms for the three fields including their interactions.  We choose an up triangle that consists of three sites of $\text{Ni}^{2+}$ (the triangular unit and site labels are shown in Fig.\ref{diagram_ugmodes} and their positions being $\mathbf{r}_{1}=0$, $\mathbf{r}_{2}=\hat{\mathbf{x}}$ and $\mathbf{r}_{3}=\frac{1}{2} \hat{\mathbf{x}}+\frac{\sqrt{3}}{2}\hat{\mathbf{y}}$) and impose inversion symmetry to obtain the normal modes. The non-trivial transformations for the up triangle are $C_3$ and $\sigma_{h} C_{2}^{\prime}$ (these transformations keep the centre of the triangle fixed, $\sigma_{h}$ is required with the $2$-fold rotation to bring the crystal field environment back to itself, details in Appendix \ref{appen_lg}).  Combining this with inversion generates all the $6$ conjugacy classes of $\text{D}_{3d}$. We can then decompose the dipole, quadrupole fields and the elastic modes into the irreducible representations (irrep) to construct the Landau-Ginzburg free energy.

\subsection{\label{sec:dipolarquadmodes}The magnetic (dipolar) and nematic (quadrupolar) modes}

Eq. \ref{Heff} the system has full SU(2) spin rotation symmetry. Hence the spin operators remain un-rotated (in spin space) under various lattice transformations while the site indices transform. Using these latter transformations, the non-trivial irreducible representations are constructed. 

\begin{table}
\begin{center}
 \begin{tabular}{ | c | c |} 
 \hline
 Irrep & Expression  \\ [0.5ex] 
 \hline
$m_{a}^{\alpha}$ & $(S_1^{\alpha}+S_2^{\alpha}+S_3^{\alpha})/\sqrt{3}$ \\ 
 \hline
$m_{e1}^{\alpha}$ & $(S_1^{\alpha}-S_2^{\alpha})/\sqrt{2}$ \\
$m_{e2}^{\alpha}$ &  $(S_1^{\alpha}+S_2^{\alpha}-2 S_3^{\alpha})/\sqrt{6}$ \\
\hline
\hline
$Q_{a}^{\alpha \beta}$ & $(Q_1^{\alpha \beta}+Q_2^{\alpha \beta}+Q_3^{\alpha \beta})/\sqrt{3}$ \\ 
 \hline
$Q_{e1}^{\alpha \beta}$ & $(Q_1^{\alpha \beta}-Q_2^{\alpha \beta})/\sqrt{2}$ \\
$Q_{e2}^{\alpha \beta}$ &  $(Q_1^{\alpha \beta}+Q_2^{\alpha \beta}-2 Q_3^{\alpha \beta})/\sqrt{6}$ \\
\hline
\end{tabular}
\end{center}
\caption{Irreducible representations for spins and quadrupoles ($\alpha, \beta=x,y,z$).}
\label{TableSpinsQuadrupoles}
\end{table}

The irreducible representations for dipoles and quadrupole on the up triangle with site labels as shown in Fig. \ref{diagram_ugmodes} are given in Table \ref{TableSpinsQuadrupoles}. 
Note that for both the fields the irreducible representations consist of a singlet ($a$) and a doublet ($e=e1,e2$). However, it is useful to note that while the dipolar field is odd under time reversal, the quadrupolar field is even. The details of the symmetry transformations of the irreps are listed in Appendix \ref{appen_lg}. All the relevant orders can be represented as different combinations of the irreducible representations. 

\subsubsection{\label{MajorMagnetic}Magnetic orders}

The relevant magnetic orders are :

\paragraph{Ferromagnetic order :} This is given by $\langle \mathbf{m}_a \rangle \neq 0 $, $\langle \mathbf{m}_{e1}\rangle = 0 $ and $\langle \mathbf{m}_{e2}\rangle = 0 $ (where bold font is used to suppress the spin index). Inverting the relations for spins in Table \ref{TableSpinsQuadrupoles}, we get : 
\begin{align}
\langle\mathbf{S}_1\rangle &=\frac{1}{6} \left(2 \sqrt{3} \langle\mathbf{m}_a\rangle+3 \sqrt{2} \langle\mathbf{m}_{e1}\rangle+\sqrt{6} \langle\mathbf{m}_{e2}\rangle \right) \nonumber\\
\langle\mathbf{S}_2\rangle &=\frac{1}{6} \left(2 \sqrt{3} \langle\mathbf{m}_a\rangle-3 \sqrt{2} \langle\mathbf{m}_{e1}\rangle+\sqrt{6} \langle\mathbf{m}_{e2}\rangle\right)\nonumber\\
\langle\mathbf{S}_3\rangle &=\frac{1}{3} \left(\sqrt{3} \langle\mathbf{m}_a\rangle-\sqrt{6} \langle\mathbf{m}_{e2}\rangle\right)
\label{invertS}
\end{align}
which is evidently a ferromagnet for only $\langle \mathbf{m}_a \rangle \neq 0 $. The ferromagnetic order also has a non-vanishing parasitic quadrupolar moment (see Appendix \ref{wavefunctions}). 

\paragraph{$120^\circ$ spiral order :} The spiral order is given by the three simultaneous conditions
\begin{align}
    &\langle \mathbf{m}_a\rangle=0,~~~~~\langle\mathbf{m}_{e1}\rangle.\langle\mathbf{m}_{e2}\rangle=0\nonumber\\
    &\langle\mathbf{m}_{e1}\rangle.\langle\mathbf{m}_{e1}\rangle=\langle\mathbf{m}_{e2}\rangle.\langle\mathbf{m}_{e2}\rangle \label{SpiralmeNorm}
\end{align}
The first condition (using Eq. \ref{invertS}) translates into zero magnetisation condition per triangle , while the second and third conditions results in the equality of magnitude of the moments at different sites, {\it i.e.}, $\langle \mathbf{S}_1\rangle.\langle\mathbf{S}_1\rangle=\langle\mathbf{S}_2\rangle.\langle\mathbf{S}_2\rangle=\langle\mathbf{S}_3\rangle.\langle\mathbf{S}_3\rangle$ and equal angle between the ordered moments, {\it i.e.}, $\langle \mathbf{S}_1\rangle.\langle\mathbf{S}_2\rangle=\langle\mathbf{S}_2\rangle.\langle\mathbf{S}_3\rangle=\langle\mathbf{S}_3\rangle.\langle\mathbf{S}_1\rangle$. From this is is fairly easy to show that the angle between any two nearest neighbour moments is $120^\circ$ such that 
\begin{eqnarray}
 &&\langle\mathbf{S}_i\rangle = \nonumber\\
 &&\frac{\sqrt{6}}{3}\bigg(\langle\mathbf{m}_{e1}\rangle \text{cos}\bigg(\mathbf{q}.\mathbf{r}_i+\frac{\pi}{6}\bigg)+\langle\mathbf{m}_{e2}\rangle \text{sin}\bigg(\mathbf{q}.\mathbf{r}_i+\frac{\pi}{6}\bigg)\bigg)\nonumber\\
 \label{LandauSpiral}
\end{eqnarray}
with $i=1,2,3$ and $\mathbf{q}=\frac{2\pi}{3}\hat{\mathbf{x}}+\frac{2\pi}{\sqrt{3}}\hat{\mathbf{y}}$. The phase factor of $\pi/6$ is due our choice of doublet modes of spins in Table. \ref{TableSpinsQuadrupoles}. Similar to the ferromagnet, the spiral order also has non-vanishing parasitic quadrupolar moments.

\subsubsection{\label{MajorNematic}Nematic orders}
For pure spin nematic orders, {the expectation value of the} magnetic moments vanishes, {{\it i.e.} $\langle {\bf S}_i\rangle=0$.} The on-site quadrupolar tensor is symmetric and can be diagonalized and using Eq.~\ref{QuadrupoleDefinition}, one can obtain the expectation values $\langle (\mathbf{S}.\hat{\mathbf{e}}_1)^2\rangle$, $\langle (\mathbf{S}.\hat{\mathbf{e}}_2)^2\rangle$ and $\langle (\mathbf{S}.\hat{\mathbf{e}}_3)^2\rangle$ along the principal axes $\hat{\mathbf{e}}_1$, $\hat{\mathbf{e}}_2$ and $\hat{\mathbf{e}}_3$ {of the quadrupole ellipsoid}.\cite{2015Kosmachev} In terms of the quadrupole ellipsoid, there are three possibilities : (1) No nematic order : $\langle (\mathbf{S}.\hat{\mathbf{e}}_1)^2\rangle=\langle (\mathbf{S}.\hat{\mathbf{e}}_2)^2\rangle=\langle (\mathbf{S}.\hat{\mathbf{e}}_3)^2\rangle$, (2) Biaxial nematic : $\langle (\mathbf{S}.\hat{\mathbf{e}}_1)^2\rangle\neq\langle (\mathbf{S}.\hat{\mathbf{e}}_2)^2\rangle\neq\langle (\mathbf{S}.\hat{\mathbf{e}}_3)^2\rangle$, and, (3) uniaxial nematic : This is obtained when $\langle (\mathbf{S}.\hat{\mathbf{e}}_i)^2\rangle=\langle (\mathbf{S}.\hat{\mathbf{e}}_j)^2\rangle\neq\langle (\mathbf{S}.\hat{\mathbf{e}}_k)^2\rangle$ for $i\neq j\neq k$. For spin-1, the first case implies a paramagnet, while the biaxial nematic is not relevant for us. Hence we focus on the uniaxial nematic where, if the unequal expectation value is greater (lesser) than the equal ones, we have a rod (disc)-like uniaxial nematic. For spin-1, only a uniaxial nematic of the disc type is allowed.\cite{penc2011spin,articleTamasToth} Therefore for the on-site quadrupolar order relevant to our calculations, the order parameter is characterised by
\begin{align}
    \langle Q^{\alpha\beta}_i\rangle=\mathcal{Q}_{i,N} \big(n^\alpha_i n^\beta_i-\delta^{\alpha\beta}/3\big)
    \label{eq:fernem}
\end{align}
with $\mathcal{Q}_{i,N}<0$ and $\mathbf{n}_i$ being the director of the nematic.

All possible three sublattice disk-like uniaxial nematic orders can be obtained from the quadrupolar irreps in Table \ref{TableSpinsQuadrupoles}. {The on-site expectation values obtained by inverting the relations are given by :}
\begin{align}
\langle Q_1^{\alpha\beta} \rangle &=\frac{1}{6} \left(2 \sqrt{3} \langle Q_a^{\alpha\beta} \rangle+3 \sqrt{2} \langle Q_{e1}^{\alpha\beta} \rangle+\sqrt{6} \langle Q_{e2}^{\alpha\beta} \rangle \right) \nonumber\\
\langle Q_2^{\alpha\beta} \rangle &=\frac{1}{6} \left(2 \sqrt{3} \langle Q_a^{\alpha\beta} \rangle-3 \sqrt{2} \langle Q_{e1}^{\alpha\beta} \rangle+\sqrt{6} \langle Q_{e2}^{\alpha\beta} \rangle\right)\nonumber\\
\langle Q_3^{\alpha\beta} \rangle &=\frac{1}{3} \left(\sqrt{3} \langle Q_a^{\alpha\beta} \rangle-\sqrt{6} \langle Q_{e2}^{\alpha\beta} \rangle\right)
\label{invertQ}
\end{align}

\paragraph{ Ferronematic order:} This is given by
\begin{align}
\langle Q_a^{\alpha\beta}\rangle = \sqrt{3} \mathcal{Q}_{FN} \big(n^{\alpha}n^{\beta}-\frac{1}{3}\delta^{\alpha\beta}\big) \label{LandauFQ},~~~~
\langle {\bf Q}_{\bf e}\rangle = 0 
\end{align}
The on-site expectation values can be obtained from Eq. \ref{invertQ} and are given by $\langle Q_1^{\alpha\beta}\rangle=\langle Q_2^{\alpha\beta}\rangle=\langle Q_3^{\alpha\beta}\rangle=\frac{1}{\sqrt{3}}~\langle Q_a^{\alpha\beta} \rangle$, 
which is clearly a ferronematic (Eq. \ref{SN1}) that is energetically favoured~\cite{PhysRevB.74.092406} when $K<0$ in Eq. \ref{Heff}. For $J>0$ and $K<0$, the ferronematic phase is stable for $|K|/J>2$ within mean-field analysis.\cite{PhysRevLett.97.087205}

\paragraph{Three sublattice nematic order:} This is given by:
\begin{align}
    \langle Q_a^{\alpha\beta} \rangle&=0 \label{3sublatticeQa}\\
\langle Q_{e1}^{\alpha\beta}\rangle &= \frac{\mathcal{Q}_{3SN}}{\sqrt{2}}(n_1^\alpha n_1^{\beta}-n_2^\alpha n_2^{\beta}) \label{3sublatticeQe1} \\
\langle Q_{e2}^{\alpha\beta}\rangle &= \frac{\mathcal{Q}_{3SN}}{\sqrt{6}}(n_1^\alpha n_1^{\beta}+n_2^\alpha n_2^{\beta}-2 n_3^\alpha n_3^{\beta})~,
\label{3sublatticeQe2}    
\end{align}
where $\mathbf{n}_1, \mathbf{n}_2, \mathbf{n}_3$ are mutually orthogonal unit vectors and $\mathcal{Q}_{3SN}<0$. Again, using Eq. \ref{invertQ}, we can see that the above conditions correspond to the three sublattice nematic order with the directors on sites $1$, $2$ and $3$ being $\mathbf{n}_1$, $\mathbf{n}_2$ and  $\mathbf{n}_3$ respectively. Such three sublattice ordering is expected to be stabilised when the sign of the biquadratic term in the spin Hamiltonian (Eq. \ref{Heff}), $K>0$. For $J>0$ and $K>0$, the three sublattice nematic is obtained for $K/J>1$.\cite{PhysRevLett.97.087205} 

\subsubsection{Coexistence of nematic order and collinear sinusoidal dipolar order} 

In passing, we point out the possibility of an interesting phase of coexisting nematic and dipole order with a three site unit-cell and hence captured  within the above formulation. This is given by $\langle \mathbf{m}_a\rangle=0~,~~~~\langle\mathbf{m}_{e1}\rangle \parallel\langle\mathbf{m}_{e2}\rangle$.
 The resultant dipolar order is collinear and sinusoidal~\cite{Kawamura_1998} with the spin configuration given by $\langle\mathbf{S}_i\rangle =\frac{\sqrt{6}}{3}\langle\mathbf{m}_{e1}\rangle \bigg(\sqrt{1+\widetilde{\lambda}^2}\bigg) \text{sin}\bigg(\mathbf{q}.\mathbf{r}_i+\frac{\pi}{6}+\phi\bigg)$, where $\mathbf{q}=\frac{2\pi}{3}\hat{\mathbf{x}}+\frac{2\pi}{\sqrt{3}}\hat{\mathbf{y}}$. The above form can be obtained from Eq. \ref{LandauSpiral} by using  $\langle\mathbf{m}_{e2}\rangle = \widetilde{\lambda} \langle\mathbf{m}_{e1}\rangle$. Thus the state corresponds to collinear sinusoidal magnetic order. Since the spins are not completely polarised it allows for (non-parasitic) nematic ordering. Using Eq. \ref{QExpectation} we find that the sinusoidal dipole ordering is accompanied by bi-axial nematic ordering with the two orthogonal principal directions of the nematic directors being along $\hat{n}=\langle\mathbf{m}_{e1}\rangle/|\langle\mathbf{m}_{e1}\rangle|$ and perpendicular to it respectively. The quadrupole moment and the magnetic moment do not have the same symmetry implying the coexistence of nematic and collinear sinusoidal dipolar order. In the present Hamiltonian we do not expect this order to be stabilised. However, they may be relevant for more generic models such as the one studied in Ref. \onlinecite{PhysRevB.106.195147}.


\subsection{\label{sec:elasticmodes}The Elastic modes}

Having discussed the dipole and the quadrupole modes of interest, we now turn to the elastic modes that they can couple to. The normal modes of a single triangle are given in Appendix \ref{Appendix:LandauNormalModes}. This consists of 
\begin{eqnarray}
\text{a singlet} : \epsilon_{a}~~~~~~~~~~~~~~~~~\text{a doublet} : (\epsilon_{e1},\epsilon_{e2})
\label{normalmodesirrep}
\end{eqnarray}

They are linearly related with the  Cartesian strain tensors, $\varepsilon_{ij}$ (where $\varepsilon_{ij}=\frac{1}{2}\left(\frac{\partial u_i}{\partial R_j}+\frac{\partial u_j}{\partial R_i}\right)$ ; $u_i$ is the $i^{th}$ component of displacement from equilibrium position, $R_i$) as   (see Appendix \ref{Appendix:LandauNormalModes})
\begin{eqnarray}
\varepsilon_{a} &=& \frac{1}{2} (\varepsilon_{xx}+\varepsilon_{yy}); \nonumber \\
\varepsilon_{e1} &=& \varepsilon_{xy}, ~~~~~~~~\varepsilon_{e2} = \frac{1}{2} (\varepsilon_{xx}-\varepsilon_{yy}) 
\label{strainsymmetry}
\end{eqnarray}

We now study their coupling with the dipole and the quadrupole modes to understand the nature of magnetoelastic response of the system.

\subsection{\label{sec:SymmetryTerms}The Landau free energy}
To write the Landau-Ginzburg free energy, we consider the long wavelength symmetry allowed terms for the dipolar, quadrupolar and the elastic modes. At the mean field level we only consider uniform terms and drop all spatially fluctuating ones. 

\subsubsection{\label{dipolequadrupole} Dipole-quadrupole free energy}
The dipole-quadrupole free energy is
\begin{equation}
    \mathcal{F}_{mQ} = \mathcal{F}_m + \mathcal{F}_Q + \mathcal{F}_{mQ}^{\text{int}} ~,
\end{equation}
where the three terms denote pure dipolar and pure quadrupolar free energies; and the interaction between the dipoles and the quadrupoles. 
\paragraph{\label{Fdipolar}Dipolar terms:}
The dipolar modes are odd under time reversal so they appear in even powers in the free energy. The dipoles have $a$ and $e$ irreps so the dipolar free energy can be written as the sum of free energy for individual modes and the interaction terms between them 
\begin{equation}
    \mathcal{F}_m = \mathcal{F}_{m,a}+\mathcal{F}_{m,e}+\mathcal{F}_{m,ae}^{int} \label{fm} ~,
\end{equation}
where upto quartic orders
\begin{equation}
\mathcal{F}_{m,a} = r_{m_a} (\mathbf{m}_{a}.\mathbf{m}_{a}) + u_{m_a} (\mathbf{m}_{a}.\mathbf{m}_{a})^2 ~, \label{fma}
\end{equation}
\begin{eqnarray}
\mathcal{F}_{m,e} &=& r_{m_e} (\mathbf{m}_{\bf e}.\mathbf{m}_{\bf e})+ u_{m_e}(\mathbf{m}_{\bf e}.\mathbf{m}_{\bf e})^2 \nonumber\\
 &+& v_{m_e}[(\mathbf{m}_{e1}.\mathbf{m}_{e2})^2-(\mathbf{m}_{e1}.\mathbf{m}_{e1})(\mathbf{m}_{e2}.\mathbf{m}_{e2})], \label{fme}
\end{eqnarray}
are the free energies of singlet and doublet modes while
\begin{align}
  \mathcal{F}_{m,ae}^{int} =& v_{m} (\mathbf{m}_{a}.\mathbf{m}_{a}) (\mathbf{m}_{\bf e}.\mathbf{m}_{\bf e})+ \Tilde{v}_{m} (m_{a}^{\alpha} m_{a}^{\beta}) (m_{e}^{\alpha}.m_{e}^{\beta})\label{fmaeint}~,
\end{align}
represents the interaction between them. 

As discussed in section \ref{MajorMagnetic}, ferromagnetic order corresponds to $\langle \mathbf{m}_a \rangle \neq 0 $ with $\langle \mathbf{m}_{\bf e}\rangle = 0$. The mean field theory leads to a continuous transition between paramagnetic (for $r_{m_a}>0$)  and ferromagnetic (for $r_{m_a}<0$) ordered state which takes place at $r_{m_a}=0$. However, for the microscopic model in Eq. \ref{Heff}, we expect that $r_{m_a}>0$. 

For the doublet mode, the continuous transition between the thermal paramagnet (for $r_{m_e}>0$) and the dipole (for $r_{m_e}<0$) ordered phase occurs at $r_{m_e}=0$ for Eq. \ref{fme}. The details of the dipole ordered phase is controlled by $v_{m_e}$.\cite{Kawamura_1998} For $v_{m_e}>0 (<0)$, the 120$^\circ$ spiral (collinear sinusoidal) state is stabilised. Note that the dipolar free energy for the doublet mode, Eq. \ref{fme}, is invariant under :
\begin{eqnarray}
\mathbf{m}_{e1} &\xrightarrow{}& \mathbf{m}_{e1}^\prime =\cos{\theta} \mathbf{m}_{e1}-\sin{\theta} \mathbf{m}_{e2} \nonumber\\
\mathbf{m}_{e2} &\xrightarrow{}& \mathbf{m}_{e2}^\prime=\pm(\sin{\theta} \mathbf{m}_{e1}+\cos{\theta} \mathbf{m}_{e2}) ~,
\label{emergentme}
\end{eqnarray}
for $\theta\in (0,2\pi]$. The origin of this enhanced symmetry is due to the enlargement of the translation symmetry from $\mathbb{Z}_3$ to $U(1)$ as can be quickly checked by writing the spin configurations in Eq. \ref{LandauSpiral}. This enhanced symmetry, absent in the microscopic model is broken down at the sixth order by the term $(\mathbf{m}_{e1}.\mathbf{m}_{e1}-\mathbf{m}_{e2}.\mathbf{m}_{e2})[(\mathbf{m}_{e1}.\mathbf{m}_{e1}-\mathbf{m}_{e2}.\mathbf{m}_{e2})^2-12(\mathbf{m}_{e1}.\mathbf{m}_{e2})^2]$.\cite{kawamura1990commensurate}

Finally the interaction term in Eq. \ref{fmaeint} leads to effective attraction (for $v_m,\tilde v_m<0$) or repulsion (for $v_m,\tilde v_m>0$) between the ferromagnetic and the spiral or the sinusoidal orders. In particular, attraction between the ferromagnetic and collinear sinusoidal order can open up a regime of ferrimagnetic order where both these order parameters are non-zero.
 
\paragraph{\label{FQuadrupolar}Quadrupolar terms:}

The quadrupolar modes are even under time reversal and can be compactly written in an SU(2) spin rotation invariant form in terms of traces over spin indices giving rise to 
\begin{equation}
    \mathcal{F}_Q = \mathcal{F}_{Q,a}+\mathcal{F}_{Q,e}+\mathcal{F}_{Q,ae}^{int}  \label{fq}~.
\end{equation}
where up to quartic orders,
\begin{equation}
\mathcal{F}_{Q,a} = r_{Q_a} \text{Tr} \mathbf{Q}_a^2 +w_{Q_a} \text{Tr} \mathbf{Q}_a^3+u_{Q_a} \text{Tr} \mathbf{Q}_a^4 \label{fqa}    
\end{equation}
\begin{align}
 &\mathcal{F}_{Q,e} = r_{Q_e} \text{Tr} (\mathbf{Q}_{e1}^2+\mathbf{Q}_{e2}^2) + s_{Q_e} \text{Tr}(\mathbf{Q}_{e2}^3-3 \mathbf{Q}_{e1}^2 \mathbf{Q}_{e2}) \nonumber \\ &+ u_{Q_e,1} \text{Tr} (\mathbf{Q}_{e1}^2+\mathbf{Q}_{e2}^2)^2 + u_{Q_e,2} [\text{Tr} (\mathbf{Q}_{e1}^2+\mathbf{Q}^2_{e2})]^2\nonumber\\
 &+u_{Q_e,3} [\text{Tr}(\mathbf{Q}_{e1}\mathbf{Q}_{e2})\text{Tr}(\mathbf{Q}_{e1}\mathbf{Q}_{e2})-\text{Tr}(\mathbf{Q}_{e1}^2)\text{Tr}(\mathbf{Q}_{e2}^2)]~.\nonumber\\ \label{fqe}   
\end{align}
for the singlet and doublet modes and
\begin{align}
 \mathcal{F}_{Q,ae}^{int} =& s_{Q_a Q_e} Q_a^{\alpha\beta}\bigg[ \bigg(Q_{e1}^{\beta\gamma}Q_{e1}^{\gamma\alpha}+Q_{e2}^{\beta\gamma}Q_{e2}^{\gamma\alpha}\bigg)\nonumber\\
 &-\frac{1}{3}\delta^{\alpha\beta}\text{Tr}(\mathbf{Q}_{e1}^2+\mathbf{Q}_{e2}^2)\bigg]  
 \label{fqaeint}~,
\end{align}
represents the leading order symmetry allowed interaction between them. On general grounds all associated transitions out of the nematic orders described by Eq. \ref{fq}, are first order (without fine-tuning) due to the presence of the third order terms. Understanding these different nematic orders is tedious due to the matrix order parameter $\mathbf{Q}$ and here we restrict ourselves to the simple nematic orders relevant to Eq. \ref{Heff}.

As discussed in section \ref{MajorNematic}, the ferronematic order corresponds to Eq. \ref{LandauFQ}. This is obtained in the regime $r_{Q_a}<0$ and $r_{Q_e}>0$ in Eq. \ref{fq} where the above solution is gotten by considering a general singlet quadrupolar tensor in its eigen-basis ($\mathbf{Q}_a = \text{diag}(l_1,l_2,-l_1-l_2)$) and extremising with respect to the eigenvalues. 

For the three sublattice nematic which is an ordering in the doublet mode, we substitute the ansatz of Eq. (\ref{3sublatticeQe1}) and (\ref{3sublatticeQe2}) in the doublet free energy Eq.(\ref{fqe}) to get,
\begin{align}
\mathcal{F}_{Q,e} =& 2 r_{Q_e} \mathcal{Q}_{3SN}^2-\frac{4}{\sqrt{6}}s_{Q_e} \mathcal{Q}_{3SN}^3\nonumber\\
&+(4 u_{Q_e,1}/3+ 4 u_{Q_e,2}-u_{Q_e,3}) \mathcal{Q}_{3SN}^4 ~.
\end{align}
For positive quartic term, we get a three sub-lattice nematic via a first order phase transition.

The interaction term between the singlet and doublet quadrupolar modes (Eq. \ref{fqaeint}) indicates that ordering of the doublet mode would give rise to a parasitic singlet quadrupole expectation value of $$Q_a^{\alpha\beta} \approx -\frac{s_{Q_a Q_e}}{2r_{Q_a}}  [(Q_{e1}^{\beta\gamma}Q_{e1}^{\gamma\alpha}+Q_{e2}^{\beta\gamma}Q_{e2}^{\gamma\alpha})-\frac{Tr(\mathbf{Q}_{\bf e}^2) \delta^{\alpha\beta}}{3}].$$
However, for the particular three sublattice nematic (Eqs. \ref{3sublatticeQe1} and \ref{3sublatticeQe2}) such parasitic ferronematic order vanishes.

\paragraph{\label{FInteractionMQ}Dipole-quadrupole interaction :}
The general form of interaction terms between dipoles and quadrupoles, due to time reversal symmetry, consists of even powers of dipoles and any powers for quadrupoles consistent with other symmetries. Therefore the lowest order term has the form $\sim m^2 Q$. This implies that pure quadrupolar ordering renormalises the mass for dipoles while dipolar ordering gives rise to parasitic quadrupole moment. The nature of the parasitic quadrupole moments is obtained by extremising the leading order dipole-quadrupole interaction :
\begin{align}
\mathcal{F}_{mQ}^{\text{int}} =& s_{m_a Q_a} Q_a^{\alpha \beta} \bigg( m_{a}^{\alpha} m_{a}^{\beta}-\frac{1}{3}(\mathbf{m}_a.\mathbf{m}_a)\delta^{\alpha\beta}\bigg)\nonumber\\
&+s_{m_e Q_a}  Q_a^{\alpha \beta} \bigg(m_{e1}^{\alpha} m_{e1}^{\beta}+m_{e2}^{\alpha} m_{e2}^{\beta}\nonumber\\
&-\frac{1}{3}(\mathbf{m}_{\bf e}.\mathbf{m}_{\bf e})\delta^{\alpha\beta}\bigg)+s_{m_e Q_e}\bigg[-2 Q_{e1}^{\alpha \beta} \bigg(\frac{m_{e1}^{\alpha}  m_{e2}^{\beta}}{2}\nonumber\\
&+\frac{m_{e2}^{\alpha}  m_{e1}^{\beta}}{2}-\frac{1}{3}(\mathbf{m}_{e1}.\mathbf{m}_{e2})\delta^{\alpha\beta}\bigg)+Q_{e2}^{\alpha \beta}\bigg(m_{e2}^{\alpha} m_{e2}^{\beta}\nonumber\\
&-m_{e1}^{\alpha} m_{e1}^{\beta}-\frac{1}{3}(\mathbf{m}_{e2}.\mathbf{m}_{e2}-\mathbf{m}_{e1}.\mathbf{m}_{e1})\delta^{\alpha\beta}\bigg)\bigg] \nonumber\\
&+u_{m_a Q_e} m_{a}^{\alpha}m_{a}^{\beta}(Q_{e1}^{\alpha\gamma}Q_{e1}^{\gamma\beta}+Q_{e2}^{\alpha\gamma}Q_{e2}^{\gamma\beta}) \nonumber\\
&+s_{m_a m_e Q_e} \bigg[Q_{e1}^{\alpha \beta} \bigg(\frac{m_{a}^{\alpha}  m_{e1}^{\beta}+m_{e1}^{\alpha}  m_{a}^{\beta}}{2}\nonumber\\
&-\frac{1}{3}(\mathbf{m}_{a}.\mathbf{m}_{e1})\delta^{\alpha\beta}\bigg)+Q_{e2}^{\alpha \beta}\bigg(\frac{m_{a}^{\alpha}  m_{e2}^{\beta}+m_{e2}^{\alpha}  m_{a}^{\beta}}{2}\nonumber\\
&-\frac{1}{3}(\mathbf{m}_{a}.\mathbf{m}_{e2})\delta^{\alpha\beta}\bigg)\bigg]
~.
\label{fmQ}
\end{align}

For the ferromagnetic ordering, this leads to \begin{equation}
     Q_{a}^{\alpha\beta} \approx -\frac{s_{m_a Q_a}}{2 r_{Q_a}}\bigg( m_{a}^{\alpha} m_{a}^{\beta}-\frac{1}{3}(\mathbf{m}_a.\mathbf{m}_a)\delta^{\alpha\beta}\bigg) ~,
\end{equation}
while for the $120^\circ$ spiral order we get
\begin{align}
    &Q_{a}^{\alpha\beta} \approx -\frac{s_{m_e Q_a}}{2 r_{Q_a}}\bigg(m_{e1}^{\alpha} m_{e1}^{\beta}+m_{e2}^{\alpha} m_{e2}^{\beta}\nonumber\\
    &~~~~~~~~~-\frac{1}{3}(\mathbf{m}_{e1}.\mathbf{m}_{e1}+\mathbf{m}_{e2}.\mathbf{m}_{e2})\delta^{\alpha\beta}\bigg)\nonumber\\
   &Q_{e1}^{\alpha\beta} \approx \frac{s_{m_e Q_e}}{r_{Q_e}}\bigg(\frac{m_{e1}^{\alpha}  m_{e2}^{\beta}+m_{e2}^{\alpha}  m_{e1}^{\beta}}{2}-\frac{1}{3}(\mathbf{m}_{e1}.\mathbf{m}_{e2})\delta^{\alpha\beta}\bigg)\nonumber\\
   &Q_{e2}^{\alpha\beta} \approx -\frac{s_{m_e Q_e}}{2 r_{Q_e}}\bigg(m_{e2}^{\alpha} m_{e2}^{\beta}-m_{e1}^{\alpha} m_{e1}^{\beta}\nonumber\\
   &~~~~~~~~~-\frac{1}{3}(\mathbf{m}_{e2}.\mathbf{m}_{e2}-\mathbf{m}_{e1}.\mathbf{m}_{e1})\delta^{\alpha\beta}\bigg)
\end{align}

The above parasitic moments for the ferromagnetic and $120^\circ$ spiral order are in accordance with what we expect from the wave functions listed in appendix \ref{wavefunctions}.
\subsubsection{\label{CouplingToMagnetic Field}Coupling with the magnetic field}
Magnetic field ($\mathbf{h} = h \hat{\mathbf{h}}$) couples to the spins via the usual Zeeman term
\begin{equation}
\mathcal{F}_{hm}=g_m \mathbf{h}.\mathbf{m}_{a}
\label{fhm}
\end{equation}
such that only the ${\bf m}_a$ mode couples linearly to the uniform magnetic field. Such linear coupling evidently favours the polarised phase along the magnetic field.

As quadrupoles are even under time reversal, the coupling takes the form:
\begin{equation}
\mathcal{F}_{hQ}=g_Q  Q_{a}^{\alpha \beta} \bigg( h^{\alpha} h^{\beta} - \frac{1}{3} \delta^{\alpha \beta} (\mathbf{h}.\mathbf{h})\bigg)
\label{fhQ}
\end{equation}
Note again, only the $Q_a$ modes couple to bilinears of the uniform magnetic field. This can be seen from the microscopic term of the form $h^\alpha h^\beta \sum_i Q_i^{\alpha\beta}$ (where $i$  is the site index) and writing it in terms of the irreducible representations in Table \ref{TableSpinsQuadrupoles}. The effect (to linear order) of small uniform magnetic field on the ferronematic and the three sublattice nematic order is summarised below.

\paragraph{Ferronematic order:}
For the ferronematic order (Eq. \ref{LandauFQ}) due to $\mathcal{F}_{hm}$ (Eq. \ref{fhm}), a uniform magnetic moment proportional to  the magnetic field develops $m_{a}^\alpha \approx -\frac{g_m h^\alpha}{2 r_{m_a}}$. Due to this uniform magnetic moment, the coupling between the ferronematic order parameter and the singlet dipolar mode in Eq. \ref{fmQ} becomes,
\begin{equation}
    \mathcal{F}_{mQ,FN}^{\text{int}} = \frac{\sqrt{3} s_{m_a Q_a} \mathcal{Q}_{FN} g_m^2 h^2}{4r_{m_a}^2} \bigg( (\hat{\mathbf{h}}.\mathbf{n})^2 - \frac{1}{3} \bigg).
    \label{fmQFN}
\end{equation}
When $g_Q\approx0$ in Eq. \ref{fhQ}, the above term alone decides whether the ferronematic directors turn perpendicular(parallel) to the magnetic field for $s_{m_a Q_a}<0$ ($s_{m_a Q_a}>0$) (since $\mathcal{Q}_{FN}<0$ for disk like ferronematic order.\cite{articleTamasToth,penc2011spin})
With $g_Q \neq 0$, $\mathcal{F}_{hQ}$ becomes,
\begin{equation}
    \mathcal{F}_{hQ, FN}= \sqrt{3} \mathcal{Q}_{FN} g_Q h^2 \bigg( (\hat{\mathbf{h}}.\mathbf{n})^2 - \frac{1}{3}  \bigg) ~,
\end{equation}
which competes with Eq. \ref{fmQFN} in deciding whether director is parallel or perpendicular to the magnetic field. Extremising the Gaussian free energy (Eq. \ref{fqa}) along with these terms, we get
\begin{align}
    \mathcal{Q}_{FN}=-\frac{\sqrt{3}(g_{Q}+s_{m_a Q_a}g_m^2/4r_{m_a}^2)}{6r_{Q_{a}}}h^2
    \label{eq_magfn}
\end{align}
while is the magnetic field induced ferronematic ordering where we have assumed for concreteness $s_{m_a Q_a},g_Q>0$.

\paragraph{Three sublattice nematic:}
For the three sublattice nematic too the uniform magnetic moment proportional to and in the direction of the magnetic field develops with $m_{a}^\alpha \approx -\frac{g_m h^\alpha}{2 r_{m_a}}$. In contrast to the ferronematic case, however, here the application of magnetic field also results in doublet dipolar modes via ${\bf m}_a$. This can be seen by extremising the $s_{m_a m_e Q_e}$ (Eq. \ref{fmQ}) :
\begin{align}
    m_{\bf e}^\alpha &\approx -\frac{s_{m_a m_e Q_e} Q_{\bf e}^{\alpha\beta} m_a^\beta}{2 r_{m_e}}=g_m\frac{s_{m_a m_e Q_e} Q_{\bf e}^{\alpha\beta}  h^\beta}{4 r_{m_e}r_{m_a}} \nonumber\\
    \label{me1me2duetoh}
\end{align}
for ${\bf e}=(e1, e2)$. It is evident that the doublet modes depend on the orientation of the magnetic field relative to the directors. When the magnetic field is along any one of the orthogonal directors of the three sublattice nematic, we see from Eq. \ref{3sublatticeQe1}, Eq. \ref{3sublatticeQe2} and Eq. \ref{me1me2duetoh} that $\mathbf{m}_{e1}$ and $\mathbf{m}_{e2}$ are also in the direction of the magnetic field. This gives rise to a collinear sinusoidal magnetization in addition to the uniform magnetization (due to $\mathbf{m}_{a}$) along the direction of the magnetic field. Considering the general case where the directors are $\mathbf{n}_1=\hat{\mathbf{x}}$, $\mathbf{n}_2=\hat{\mathbf{y}}$ and $\mathbf{n}_3=\hat{\mathbf{z}}$ and the magnetic field is at a general inclination to the directors, $\mathbf{h}=h(\sin{\theta}\cos{\phi}\hat{\mathbf{x}} +\sin{\theta}\sin{\phi}\hat{\mathbf{y}}+\cos{\theta}\hat{\mathbf{z}})$, using Eq. \ref{invertS} we see that the resulting doublet and singlet magnetisations correspond to,
$\langle\mathbf{S}_1\rangle=(a_{m} \sin{\theta}\cos{\phi}\hat{\mathbf{x}} +b_{m}\sin{\theta}\sin{\phi}\hat{\mathbf{y}}+b_{m}\cos{\theta}\hat{\mathbf{z}})$, $\langle\mathbf{S}_2\rangle=(b_{m} \sin{\theta}\cos{\phi}\hat{\mathbf{x}} +a_{m}\sin{\theta}\sin{\phi}\hat{\mathbf{y}}+b_{m}\cos{\theta}\hat{\mathbf{z}})$ and $\langle\mathbf{S}_3\rangle=(b_{m} \sin{\theta}\cos{\phi}\hat{\mathbf{x}} +b_{m}\sin{\theta}\sin{\phi}\hat{\mathbf{y}}+a_{m}\cos{\theta}\hat{\mathbf{z}})$ where $a_m=\frac{g_m h}{3}\left(-\frac{\sqrt{3}}{2 r_{m_a}}+\frac{2s_{m_a m_e Q_e} Q_{3SN}}{4 r_{m_e}r_{m_a}}\right)$ and $b_m=\frac{g_m h}{3}\left(-\frac{\sqrt{3}}{2 r_{m_a}}-\frac{s_{m_a m_e Q_e} Q_{3SN}}{4 r_{m_e}r_{m_a}}\right)$. The resultant phase is a combination of uniform polarization along the field as well as sub-lattice dependent polarisation along the three orthogonal direction. 

\subsubsection{\label{Elastic}Elastic term and coupling of strains to dipoles and quadrupoles}

Finally, we turn to the contributions to the free energy due to elastic fields as well as spin-phonon coupling.
\paragraph{Elastic energy:}
The harmonic elastic energy for the singlet and the doublet strain modes is
\begin{equation}
\mathcal{F}_{\text{elastic}}=\frac{1}{2} (c_{a} \varepsilon_{a}^{2}+ c_{e} \varepsilon_{e}^{2}) \label{felastic}    ~,
\end{equation}
where $c_a$ and $c_e$ are two independent elastic constants.

\paragraph{Coupling between dipoles, quadrupoles and strains :} The coupling term between dipoles and strain fields is given by
\begin{equation}
    \mathcal{F}_{\varepsilon m} = \mathcal{F}_{\varepsilon ma}+\mathcal{F}_{\varepsilon me} \label{fEm} ~,
\end{equation}
where
\begin{eqnarray}
    \mathcal{F}_{\varepsilon ma} &=& [\mu_{a1}\varepsilon_{a}+\frac{\mu_{a}}{2} \varepsilon_{a}^{2}+\frac{\mu_{e}}{2} \varepsilon_{e}^{2}]p_1 (\mathbf{m}_{a}.\mathbf{m}_{a}) \label{fEma}\\
    \mathcal{F}_{\varepsilon me} &=&  [\mu_{a1}\varepsilon_{a}+\frac{\mu_{a}}{2} \varepsilon_{a}^{2}+\frac{\mu_{e}}{2} \varepsilon_{e}^{2}]p_2 (\mathbf{m}_{\bf e}.\mathbf{m}_{\bf e})
    \label{fEme}
\end{eqnarray}
denotes the coupling between the singlet and doublet dipolar modes.

Similarly, the coupling term between quadrupoles and strains is given by
\begin{equation}
    \mathcal{F}_{\varepsilon Q} = \mathcal{F}_{\varepsilon Qa}+\mathcal{F}_{\varepsilon Qe} \label{fEq} ~,
\end{equation}
where
\begin{eqnarray}
\mathcal{F}_{\varepsilon Qa} &=&  [\Tilde{\mu}_{a1}\varepsilon_{a}+\frac{\Tilde{\mu}_{a}}{2} \varepsilon_{a}^{2}+\frac{\Tilde{\mu}_{e}}{2} \varepsilon_{e}^{2}]p_3 \text{Tr}\mathbf{Q}_{a}^2\label{fEqa}\\
\mathcal{F}_{\varepsilon Qe} &=&  [\Tilde{\mu}_{a1}\varepsilon_{a}+\frac{\Tilde{\mu}_{a}}{2} \varepsilon_{a}^{2}+\frac{\Tilde{\mu}_{e}}{2} \varepsilon_{e}^{2}]p_4 \text{Tr}(\mathbf{Q}_{\bf e}.\mathbf{Q}_{\bf e})
\label{fEqe}
\end{eqnarray}

In both the cases of dipolas and quadrupoles, the free energy allows for linear coupling with the singlet strain field with bilinear of the dipole/quadrupole fields. This would lead to lattice distortions upon dipolar or quadrupolar ordering while the quadratic terms (in elastic fields) lead to renormalisation of elastic constants and hence the sound speed. 

\section{\label{Observables}Fractional change in sound speed and fractional change in length}

\subsection{Fractional change in length}
Fractional change in length along direction $\hat{r}\equiv(\cos\theta,\sin\theta)$ (where $\theta$ is the angle with respect to the Cartesian $x$-axis in Fig. \ref{diagram_ugmodes}) can be obtained from the Cartesian strain fields $\varepsilon_{ij}$ :
\begin{align}
\bigg(\frac{\Delta L}{L}\bigg)_{\hat{r}} &= \sum_{ij} \varepsilon_{ij}\hat{r}_i \hat{r}_j=\varepsilon_a +\varepsilon_{e2}\cos(2\theta)+\varepsilon_{e1}\sin(2\theta).
\label{fracL}
\end{align}
Due to the linear coupling term between the uniform strain field, $\varepsilon_a$ and the dipole/nematic bilinears (Eqs. \ref{fEma}, \ref{fEme}, \ref{fEqa} and \ref{fEqe}), this leads to isotropic magnetostriction of the triangular lattice given by

\begin{eqnarray}
\bigg(\frac{\Delta L}{L}\bigg)_{\hat{r}}=\varepsilon_{a} &\approx& -\frac{1}{c_a} \bigg[ \mu_{a1} \bigg( p_1 (\mathbf{m}_{a}.\mathbf{m}_{a}) + p_2 (\mathbf{m}_{\bf e}.\mathbf{m}_{\bf e}) \bigg) \nonumber\\
&+& \Tilde{\mu}_{a1} \bigg( p_3 \text{Tr}\mathbf{Q}_{a}^2 + p_4 \text{Tr}(\mathbf{Q}_{\bf e}.\mathbf{Q}_{\bf e}) \bigg) \bigg]
\end{eqnarray}

Focussing on the $120^\circ$ spiral order, Eq. \ref{fracL} reduces to
\begin{align}
    \bigg(\frac{\Delta L}{L}\bigg)_{\hat{r}}=-\frac{\mu_{a1}p_2}{c_a}  (\mathbf{m}_{\bf e}.\mathbf{m}_{\bf e})
\end{align}
such that below the critical point ($r_{m_e}<0$ in Eq. \ref{fme}) when ${\bf m}_e\sim \sqrt{-r_{m_e}/u_{m_e}}$,
\begin{align}
    \bigg(\frac{\Delta L}{L}\bigg)_{\hat{r}}\propto |r_{m_e}|
\end{align}
as expected from the lowest symmetry allowed coupling. Thus for a thermal phase transition where $r_{m_e}\propto (T-T_c)$, one expects a linear turning on of the magnetiostrictive distortion with measurable consequence for thermal expansion experiments.

A more startling effect occurs across a spin nematic transition. In particular for a ferronematic, the above expression reduces to
\begin{align}
    \bigg(\frac{\Delta L}{L}\bigg)_{\hat{r}}=-\frac{\mu_{a1}p_3}{c_a}  Tr{\bf Q}_a^2=-\frac{2\mu_{a1}p_3}{c_a}\mathcal{Q}_{FN}^2
\end{align}
such that across the discontinuous nematic transition, there is a simultaneous jump in the lattice volume.

On turning on the magnetic field inside either the spiral or the ferronematic, there are added contributions due to the Zeeman term (Eq. \ref{fhm}) resulting in non-zero $m_a^\alpha=-\frac{g_mh^\alpha}{2r_{m_{a}}}$ and $\mathcal{Q}_{FN}$ (Eq. \ref{eq_magfn}) such that we have
\begin{align}
    \bigg(\frac{\Delta L}{L}\bigg)_{\hat{r}}=&-\frac{\mu_{a1}p_2}{c_a}  (\mathbf{m}_{\bf e}.\mathbf{m}_{\bf e})-\frac{\mu_{a1}p_1g_m^2}{4c_ar_{m_{a}}^2}h^2\nonumber\\
    &-\frac{\tilde\mu_{a1}p_3 (g_{Q}+ s_{m_a Q_a} g_m^2/4r_{m_a}^2)^2}{6r_{Q_{a}}^2c_a}~~h^4
\end{align}
for the spiral and 
\begin{align}
    \bigg(\frac{\Delta L}{L}\bigg)_{\hat{r}}=&-\frac{2\mu_{a1}p_3}{c_a}\mathcal{Q}_{FN}^2-\frac{\mu_{a1}p_1g_m^2}{4c_ar_{m_{a}}^2}h^2\nonumber\\
    &-\frac{\tilde\mu_{a1}p_3 (g_{Q}+ s_{m_a Q_a} g_m^2/4r_{m_a}^2)^2}{6r_{Q_{a}}^2c_a}~~h^4
\end{align}
for the ferronematic.

While both the forms predict a similar mixture of $h^2$ and $h^4$ dependence, one expects that near the ferronematic phase transition $r_{Q_a}\rightarrow 0$ such that the $h^4$ dependence becomes more pronounced on approaching the ferronematic phase transition. Similar results hold for the three sublattice nematic.
\subsection{Fractional change in sound speed}
The fractional change in sound speed, $v$, is related to the change in elastic constants, $c$, (Eq. \ref{felastic}) as : 
\begin{equation}
\frac{\Delta v}{v} \propto \frac{\Delta c}{c} ~,
\label{fracv}
\end{equation}

The component of the elastic tensor depends on the propagation direction and polarization of the sound and hence is generically a linear combination of $c_a$ and $c_e$.

The renormalisation of these two elastic constants are due to coupling with the dipole and nematic orders are readily obtained from Eqs. \ref{fEm} and \ref{fEq} as :
\begin{eqnarray}
\frac{\Delta c_a}{c_a} &=& \frac{1}{c_a} \bigg[ \mu_{a} \bigg( p_1 (\mathbf{m}_{a}.\mathbf{m}_{a}) + p_2 (\mathbf{m}_{\bf e}.\mathbf{m}_{\bf e}) \bigg) \nonumber\\
&+& \Tilde{\mu}_{a} \bigg( p_3 \text{Tr}\mathbf{Q}_{a}^2 + p_4 \text{Tr}(\mathbf{Q}_{\bf e}.\mathbf{Q}_{\bf e}) \bigg) \bigg] \nonumber\\
\frac{\Delta c_e}{c_e} &=& \frac{1}{c_e} \bigg[ \mu_{e} \bigg( p_1 (\mathbf{m}_{a}.\mathbf{m}_{a}) + p_2 (\mathbf{m}_{\bf e}.\mathbf{m}_{\bf e}) \bigg) \nonumber\\
&+& \Tilde{\mu}_{e} \bigg( p_3 \text{Tr}\mathbf{Q}_{a}^2 + p_4 \text{Tr}(\mathbf{Q}_{\bf e}.\mathbf{Q}_{\bf e}) \bigg) \bigg]
\end{eqnarray}

Therefore similar to the case of fractional change in length, the fraction change in sound speed would be continuous across the $120^\circ$ spiral ordering transition and discontinuous across the nematic transitions. The magnetic field dependence below the ordering transitions will have a similar dependence.

Observing jumps in fractional change in length and fractional change in sound speed in the absence of any magnetization, along with the enhanced $h^4$ magnetic field dependence  would strongly favour the case for spin-nematic orders.

\section{Summary and Outlook}

In this work, we have studied the possible elastic signatures of a spin-nematic, possibly realised in the triangular lattice compound NiGa$_2$S$_4$. Inspired by recent Raman scattering experiments~\cite{PhysRevLett.125.197201} which indicate substantial spin-phonon coupling in the material, and the general expectation that the time reversal symmetric spin-nematic order parameter can linearly couple to the elastic degrees of freedom, we show that-- (1) inside such a spin-nematic phase, the interaction between the nematic Goldstone boson in a ferronematic phase and the phonons lead to a powerlaw dependence of the fractional change in sound speed, {\it i.e.}, $\Delta v/v\propto T^3$, (2) across the spin-nematic phase transition, there is a finite jump in both magnetostriction as well as $\Delta v/v$ stemming from the discontinuous nature of the nematic transition, and (3) possible enhanced $h^4$ dependence of the magnetostriction and $\Delta v/v$ on the magnetic field just below the nematic transition. These signatures, along with thermodynamic measurements such as low temperature powerlaw specific heat~\cite{Nakatsuji1697,doi:10.1143/JPSJ.75.083701,PhysRevB.74.092406,PhysRevLett.97.087205,PhysRevB.79.214436} and absence of long range spin correlations provides important steps in characterising the spin-nematic phase in particular and paves a way for concrete experimental signatures for higher-moment magnetic orderings that have proved quite challenging in spite of several candidate materials. It is useful to note that light scattering can also act as a complementary spectroscopic probe for spin-nematic as shown in Ref. \onlinecite{PhysRevB.84.184424}.

In context of NiGa$_2$S$_4$, we have used the minimal nearest neighbour bilinear-biquadratic spin-1 Hamiltonian (Eq. \ref{Heff}) to understand the the low temperature elastic signatures. In the actual material, in addition to nearest neighbour bilinear and biquadratic terms, a third neighbour antiferromagnetic bilinear spin  interaction appears to be relevant to understand the detailed physics. Here, however, we have  neglected such terms as we are interested in the elastic response of the spin-nematic where the effect of such terms are expected to be secondary and as far as the temperature dependence is concerned. We further note that in addition to spin-rotation invariant terms, the microscopics of NiGa$_2$S$_4$ can also admit small single-ion anisotropies as well as DM interactions. While the former can pin the nematic director,\cite{PhysRevB.74.092406} the latter can pin the chirality of the spiral state for the DM vector pointing perpendicular to the plane (see Appendix \ref{sec:AppendixMFTPhases}). Therefore the former can gap out the nematic Goldstone modes and change the powerlaw dependence to an exponential damping for $\Delta v/v$. However, since such effects have not been measured in the low temperature specific heat,\cite{Nakatsuji1697,doi:10.1143/JPSJ.75.083701,PhysRevB.74.092406,PhysRevLett.97.087205,PhysRevB.79.214436} we neglect such single-ion anisotropies. 

Several studies on ultrasound properties on candidates for multipolar order in SOC coupled systems~\cite{patri2019unveiling} or in-field spin-nematic~\cite{PhysRevResearch.1.033065} exists. We hope our results will generate interest in probing ultrasound renormalisation in NiGa$_2$S$_4$ which is a rare spin-nematic candidate in a system described by a spin rotation symmetric spin Hamiltonian to a very good approximation. 

\acknowledgments
The authors thank R. Moessner, S. Nakatsuji, N. Drichko and S. Zherlitsyn for discussion. SB acknowledges adjunct fellow program at SNBNCBS, Kolkata for hospitality. The authors  acknowledge funding from Max Planck Partner group Grant at ICTS, Swarna Jayanti fellowship grant of SERB-DST (India) Grant No. SB/SJF/2021-22/12 and the Department of Atomic Energy, Government of India, under Project No. RTI4001.
\appendix
\section{\label{wavefunctions} Wave function for various spin-1 orders}

The mean-field product wave functions for spin-1 magnets for the dipole and nematic orders \cite{PhysRevLett.97.087205,penc2011spin,articleTamasToth} relevant to this work is best understood in the basis :
\begin{eqnarray}
\ket{x}=i\frac{\ket{1}-\ket{\bar{1}}}{\sqrt{2}},~
\ket{y}=\frac{\ket{1}+\ket{\bar{1}}}{\sqrt{2}},~
\ket{z}=-i\ket{0}.
\label{xyzbasis}
\end{eqnarray}
The on-site spin and the quadrupole operators are
\begin{align}
S^{\Bar{\mu}} = -i\sum_{\Bar{\nu}\Bar{\lambda}}\epsilon_{\Bar{\mu}\Bar{\nu}\Bar{\lambda}}\ket{\Bar{\nu}}\bra{\Bar{\lambda}};~
Q^{\Bar{\mu}\Bar{\nu}} = \frac{1}{3}\delta^{\Bar{\mu}\Bar{\nu}}-\frac{\ket{\Bar{\nu}}\bra{\Bar{\mu}}}{2}-\frac{\ket{\Bar{\mu}}\bra{\Bar{\nu}}}{2}
\end{align}
where $\epsilon_{\Bar{\mu}\Bar{\nu}\Bar{\lambda}}$ denotes the Levi Civita tensor and $\Bar{\mu}$, $\Bar{\nu}$, $\Bar{\lambda}=x,y,z$. The most general single spin-1 state is 
\begin{equation}    \ket{\mathbf{w}}=\sum_{\Bar{\mu}}w_{\Bar{\mu}}\ket{\Bar{\mu}} ~,
\label{generalstate}
\end{equation}
where $w_{\Bar{\mu}}$ are the three components of a complex vector 
\begin{equation}
    \mathbf{w}=\mathbf{u}+i \mathbf{v}
\end{equation}
with ${\bf u}, {\bf v} \in \mathcal{R}^3$ and  $\mathbf{u}\cdot\mathbf{u}+\mathbf{v}\cdot\mathbf{v}=1$ with the overall phase fixed by  $\mathbf{u}\cdot\mathbf{v}=0$. The expectation of spin and quadrupole operators are :
\begin{eqnarray}
\bra{\mathbf{w}}\mathbf{S}\ket{\mathbf{w}} &=& 2 \mathbf{u} \times \mathbf{v} \label{SpinExpectation}\\
\bra{\mathbf{w}}Q^{\Bar{\mu}\Bar{\nu}}\ket{\mathbf{w}} &=& \frac{1}{3}\delta^{\Bar{\mu}\Bar{\nu}}-u^{\Bar{\mu}}u^{\Bar{\nu}}-v^{\Bar{\mu}}v^{\Bar{\nu}} \label{QExpectation}~.
\end{eqnarray}
The mean field wave function for different magnetic and nematic ordered states are direct products of the onsite wave functions, $\ket{\psi}_{MF}=\otimes_{i}\ket{\mathbf{w}_i}$.   
\paragraph*{Ferromagnetic order:}
The ferromagnetic order in the $\hat{\mathbf{z}}$ direction is given by
\begin{align}
    \mathbf{u}_i &=\frac{1}{\sqrt{2}}(\text{sin}\theta\text{ } \hat{\mathbf{x}}+\text{cos}\theta\text{ }\hat{\mathbf{y}}),~~
    \mathbf{v}_i =\frac{-1}{\sqrt{2}}(\text{cos}\theta\text{ } \hat{\mathbf{x}}-\text{sin}\theta\text{ }\hat{\mathbf{y}}),
    \label{wfmagnetic}
\end{align}
such that the spin expectation value (Eq. \ref{SpinExpectation}) is given by
\begin{equation}
    \bra{\mathbf{w}}\mathbf{S}\ket{\mathbf{w}} = \text{ }\hat{\mathbf{z}}
\end{equation}
which characterises the fully polarised ferromagnetic state. The on-site parasitic quadrupole order is
\begin{equation}
\bra{\mathbf{w}}Q^{\Bar{\mu}\Bar{\nu}}\ket{\mathbf{w}} = \begin{pmatrix}
    -\frac{1}{6} & 0 & 0 \\
    0 & -\frac{1}{6} & 0 \\
    0 & 0 & \frac{1}{3} \\
    \end{pmatrix} ~,
    \label{QFullyPolarized}
\end{equation}
in the $ \{\hat{\mathbf{x}},\hat{\mathbf{y}},\hat{\mathbf{z}}\} $ basis and corresponds to uniaxial parasitic ferronematic order with director along $\hat{\mathbf{z}}$. 

\paragraph*{$120^\circ$ spiral order:}
For a $120^\circ$ spiral order : 
\begin{align}
    \mathbf{u}_1 &= \frac{1}{\sqrt{2}} \hat{\mathbf{z}} \text{  ,   } \mathbf{v}_1 = -\frac{1}{\sqrt{2}} \hat{\mathbf{y}} \nonumber\\
    \mathbf{u}_2 &= \frac{1}{\sqrt{2}} \hat{\mathbf{z}} \text{  ,   } \mathbf{v}_2 = -\frac{1}{\sqrt{2}}\bigg(\frac{\sqrt{3}}{2} \hat{\mathbf{x}}-\frac{1}{2}\hat{\mathbf{y}}\bigg) \nonumber\\
    \mathbf{u}_3 &= \frac{1}{\sqrt{2}} \hat{\mathbf{z}} \text{  ,   } \mathbf{v}_3 = -\frac{1}{\sqrt{2}}\bigg(-\frac{\sqrt{3}}{2} \hat{\mathbf{x}}-\frac{1}{2}\hat{\mathbf{y}}\bigg)
\end{align}
such that the spin expectation value is given by Eq. \ref{spiral1} and corresponds to Fig. \ref{PhaseDiagramFig}. Again, each on-site parasitic quadrupolar moment would be uniaxial with the director being in the direction of the respective magnetic moment.

Pure nematic orders require $\bra{\mathbf{w}}\mathbf{S}\ket{\mathbf{w}} = 2 \mathbf{u} \times \mathbf{v} = 0 $ in addition to $\mathbf{u}\cdot\mathbf{v}=0$ which correspond to either $\mathbf{v} = 0$ or $\mathbf{u} = 0$ {\it i.e.}, $\mathbf{w}$ is real or imaginary. Considering $\mathbf{v} = 0$, the expectation value of the quadrupole operator (Eq. \ref{QExpectation}) is,
\begin{equation}
\bra{\mathbf{w}}Q^{\Bar{\mu}\Bar{\nu}}\ket{\mathbf{w}} = \frac{1}{3}\delta^{\Bar{\mu}\Bar{\nu}}-u^{\Bar{\mu}}u^{\Bar{\nu}} ~,
\end{equation}
which is a uniaxial nematic with the director along $\mathbf{u}$. 
\paragraph*{Ferronematic order:}
All sites have  
\begin{align}
\mathbf{u}&=\hat{\mathbf{n}},~~~~\mathbf{v}=0
\end{align}
with $\hat{\mathbf{n}}$ being the director  (Fig.\ref{diagram_ugmodes}). 
\paragraph*{Three sublattice nematic order:}
The three sites of the triangular unit (see Fig. \ref{diagram_ugmodes})) have :
\begin{align}
    \mathbf{u}_1&=\hat{\mathbf{n}}_1 \text{ , } \mathbf{v}_1=0 \nonumber\\
    \mathbf{u}_2&=\hat{\mathbf{n}}_2 \text{ , } \mathbf{v}_2=0 \nonumber\\
    \mathbf{u}_3&=\hat{\mathbf{n}}_3 \text{ , } \mathbf{v}_3=0 ~
    \label{Eq_3snop}
\end{align}
where $\hat{\mathbf{n}}_1$ , $\hat{\mathbf{n}}_2$ and $\hat{\mathbf{n}}_3$ are mutually orthogonal.

\paragraph*{Co-existing nematic order and collinear sinusoidal dipolar order:} The spins for a three sublattice collinear sinusoidal dipolar order (with co-existing nematic order discussed in Section \ref{MajorNematic}) can be expressed as $
    \langle \mathbf{S}_i \rangle = \text{sin}\eta_i \text{ } \hat{m}$ (with $\eta_i=\mathbf{q}.\mathbf{r}_i+\pi/6+\phi$).  Using Eq. \ref{wfmagnetic}, the wave functions with the above order are :
\begin{align}
    \mathbf{u}_i &=\text{cos}(\eta_i/2)(\text{sin}\theta_i\text{ } \hat{\mathbf{m}}_{\perp_1}+\text{cos}\theta_i\text{ }\hat{\mathbf{m}}_{\perp_2})\nonumber\\
    \mathbf{v}_i &=-\text{sin}(\eta_i/2)(\text{cos}\theta_i\text{ } \hat{\mathbf{m}}_{\perp_1}-\text{sin}\theta_i\text{ }\hat{\mathbf{m}}_{\perp_2})
    \label{WFCollinearSinusoidal}
\end{align}
where $\hat{\mathbf{m}}_{\perp_1}\times\hat{\mathbf{m}}_{\perp_2}=\hat{\mathbf{m}}$, $i=1,2,3$ and $ \eta_2=\eta_1+2\pi/3$ ,  $\eta_3=\eta_1+4\pi/3$. 
Such an order would feature sites having partially polarized spins. Unlike quadrupole moments for fully polarized spins (Eq. \ref{QFullyPolarized}), for the case of partially polarized spins one can see that the expectation values of the on-site quadrupolar moment (using Eq. \ref{QExpectation}) would depend on the choice of $\theta_i$ for the site and the quadrupole moment would be bi-axial - with the eigenvalues being $\{1/3,1/3-|\mathbf{u}_i|^2,1/3-|\mathbf{v}_i|^2\}$ for the eigenvectors being $\{\hat{\mathbf{m}},\hat{\mathbf{u}}_i,\hat{\mathbf{v}}_i\}$ respectively. 

\section{\label{sec:AppendixMFTPhases} Mean field T=0 phase diagram and DM term}

\begin{figure}
\begin{center}
\includegraphics[width=\linewidth,keepaspectratio]{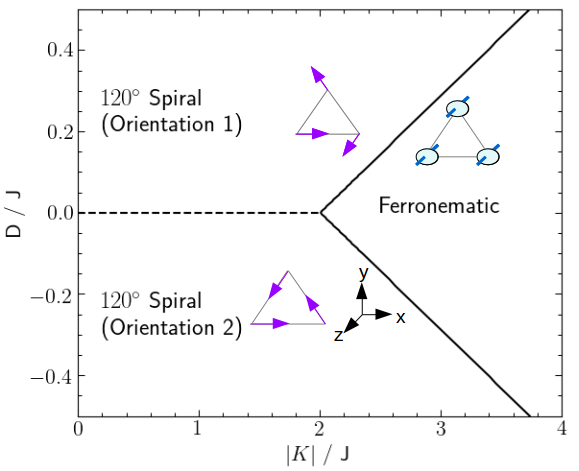}
\end{center}
\caption{Mean field $T=0$ phase for bi-linear biquadratic spin-1 ($J>0$ and $K<0$) Hamiltonian with the DM interaction. Dashed line denotes first order transition and continuous lines are for continuous transition. Orientation 1 has spiral $\mathbf{q}=-\frac{2\pi}{3}\hat{\mathbf{x}}-\frac{2\pi}{\sqrt{3}}\hat{\mathbf{y}}$, Orientation 2 has spiral $\mathbf{q}=\frac{2\pi}{3}\hat{\mathbf{x}}+\frac{2\pi}{\sqrt{3}}\hat{\mathbf{y}}$ (lattice constant taken to be unity).}
\label{PhaseDiagramFig} 
\end{figure}

\begin{figure}
\begin{center}
	\includegraphics[scale=0.25]{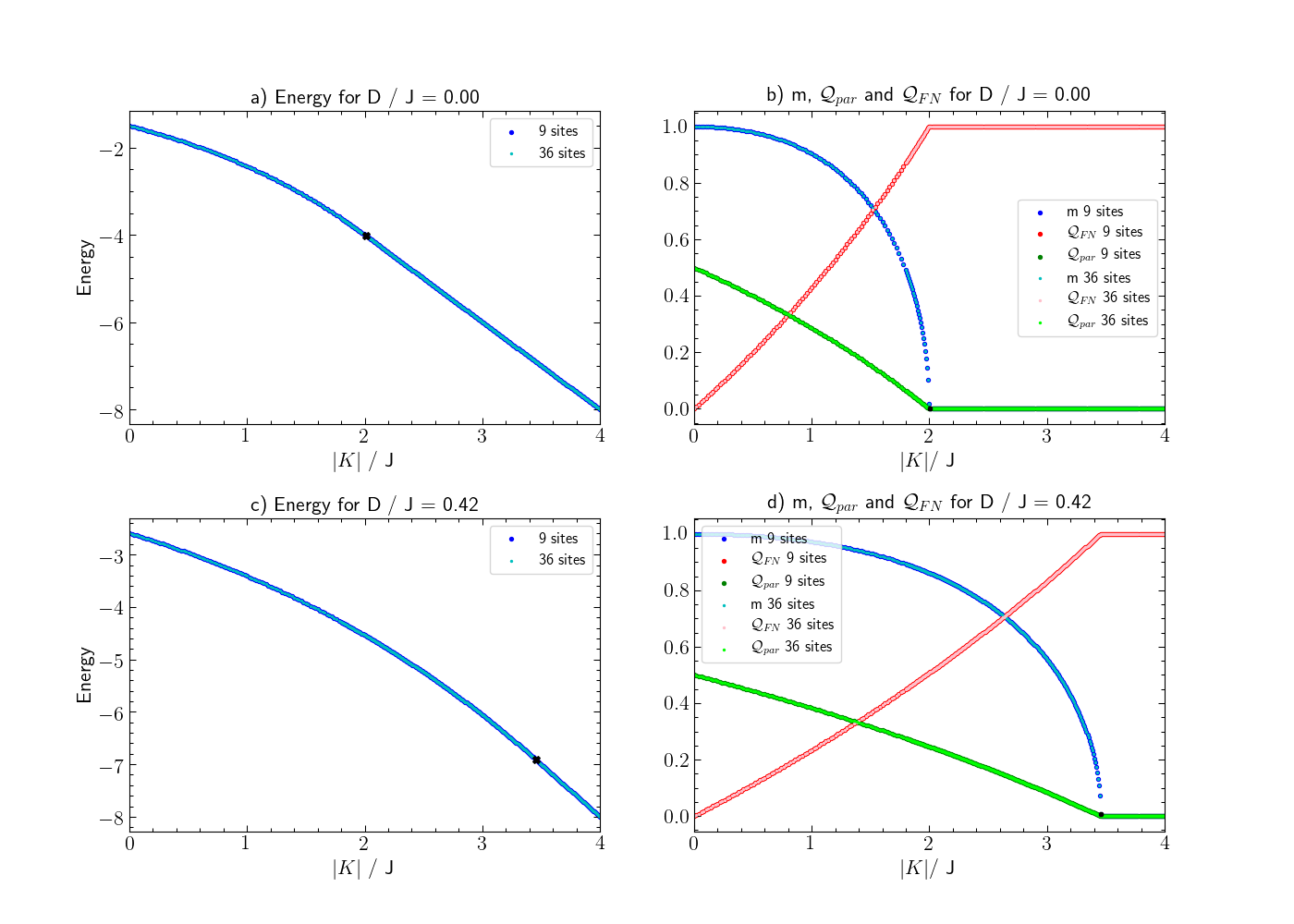}
\end{center}	
\caption{Results of the self consistency calculations (for $J>0$ and $K<0$) : a),b): Energy and order parameters for $D=0$, c),d): Energy and order parameters for $D/J=0.42$}
\label{combined} 
\end{figure}

The mean field phase diagram of Eq. \ref{Heff} was obtained in Ref. \onlinecite{PhysRevB.74.092406}. On adding to it,  a DM term given by $H_{DM} = \mathbf{D}_{ij}\cdot(\mathbf{S}_i \times \mathbf{S}_j)$, where $\mathbf{D}_{ij}=-\mathbf{D}_{ji}$ and $\mathbf{D}_{ij}=D\hat{\mathbf{z}}$ vector pointing out of the triangular lattice plane, the mean field phase diagram is given by Fig. \ref{PhaseDiagramFig} for $J>0$ and $K<0$. The main role of the DM term is to lift the degeneracy of the clockwise and anti-clockwise spirals. The different order parameters are plotted in Fig. \ref{combined}.

\section{Linear Spin-nematic  wave theory}
\label{sec:AppendixSW}

 \subsection{\label{sec:AppendixSWFN}The ferronematic}

As discussed in Sec. \ref{sec:dvbyvFerronematic}, the Goldstone modes out of a ferronematic with director along the $\hat{\bf z}$ direction can be captured by defining two bosons at each site $i$, $b_{i1}^{\dagger}$ and $b_{i2}^{\dagger}$. For a faithful description of the spin-1 Hilbert space, the bosons are subject to the constraints -- $b_{i1}^{\dagger} b_{i1}^{\dagger}=b_{i1} b_{i1}=b_{i2}^{\dagger} b_{i2}^{\dagger}=b_{i2} b_{i2}=b_{i1}^{\dagger} b_{i2}^{\dagger}=b_{i1}b_{i2}=b_{i1}b_{i2}^{\dagger}=b_{i2}b_{i1}^{\dagger}=0$ and $b_{i2}b_{i2}^{\dagger}=b_{i1}b_{i1}^{\dagger}=1-b_{i1}^{\dagger} b_{i1}-b_{i2}^{\dagger} b_{i2}$. The spin operators can be written in terms of these bosons as
\begin{align}
S_{i}^{+} = \sqrt{2} [b_{i1}^{\dagger}+b_{i2}],~S_{i}^{-} = \sqrt{2} [b_{i1}+b_{i2}^{\dagger}],~S_{i}^{z} = n'_{i1}-n'_{i2},
\label{swspin}
\end{align} 
where, $n'_{i1(2)}=b_{i1(2)}^\dagger b_{i1(2)}$. It is useful to note that the above two bosons are related to the three bosons $a_x, a_y, a_z$ of the SU(3) flavour wave theory~\cite{PhysRevLett.97.087205} as : $b_{1}=(-a_{x}+ i  a_{y})/\sqrt{2}$ and $b_{2}=(a_{x}+ i a_{y})/\sqrt{2}$ while $a_z$ is condensed. Using Eq. \ref{swspin} in Eq. \ref{Heff} and using the constraints, we get 
\begin{eqnarray}
H_{sp, FN} &=& 6 K_0 N - 6 K_0 \sum_{i,\mu'} n'_{i\mu'} +2 J_0\sum_{\langle ij\rangle}(b_{i1}^{\dagger} b_{j1}+b_{i2}^{\dagger} b_{j2})\nonumber\\&&+2(J_0-K_0)\sum_{\langle ij\rangle}(b_{i1}^{\dagger} b_{j2}^{\dagger}+b_{i1} b_{j2})
\end{eqnarray}
where $\mu'=1,2$. The quadratic spin-nematic wave Hamiltonian above can be diagonalized via Fourier transform followed by Bogoliubov transformation, 
\begin{eqnarray}
\left(\begin{matrix} b_{\mathbf{k},1}\\b_{-\mathbf{k},2}^{\dagger} \end{matrix}\right)
&=& \mathcal{R}_{\mathbf{k}} \psi_{\mathbf{k}} ~, 
\label{BogoDefn}
\end{eqnarray}
with
\begin{eqnarray}
\mathcal{R}_{\mathbf{k}} &=&\left(\begin{matrix} u_{\mathbf{k}} & v_{\mathbf{k}} \\ v_{\mathbf{k}} & u_{\mathbf{k}}\end{matrix}\right) \text{ and  } \psi_{\mathbf{k}} =\left(\begin{matrix}d_{\mathbf{k},1}\\d_{-\mathbf{k},2}^{\dagger}\end{matrix}\right)
\label{Rdefn}
\end{eqnarray}
such that $u_{\mathbf{k}}=\text{cosh}\Bar{\theta}_{\mathbf{k}}$, 
$v_{\mathbf{k}} = \text{sinh}\Bar{\theta}_{\mathbf{k}}$ and $ \text{tanh} 2\Bar{\theta}_{\mathbf{k}} = \frac{(K_0-J_0)\gamma_{\mathbf{k}}}{J_0 \gamma_{\mathbf{k}}-K_0}$. The diagonalized Hamiltonian and dispersion are given in Eq. \ref{HspFN} and Eq. \ref{HspFNdispersion} respectively. 

\subsection{\label{sec:AppendixSW3SN}The  three sublattice nematic}
Spin-nematic wave theory for the three sublattice nematic,\cite{doi:10.1143/JPSJ.75.083701} as mentioned in the main text, requires two bosons at every sublattice, $\tilde\alpha_i$ and $\tilde\beta_i$, that account for the deviations about the respective nematic director. In terms of these bosons, the spin operators at sublattice $3$ (say) (see Fig. \ref{diagram_ugmodes}) are given by
\begin{eqnarray}
S^x_{\mathbf{\widetilde{R}},3} &=& \widetilde{\alpha}^\dagger_{\mathbf{\widetilde{R}},3} + \widetilde{\alpha}_{\mathbf{\widetilde{R}},3},~~~
S^y_{\mathbf{\widetilde{R}},3} = \widetilde{\beta}^\dagger_{\mathbf{\widetilde{R}},3} + \widetilde{\beta}_{\mathbf{\widetilde{R}},3} \nonumber\\
S^z_{\mathbf{\widetilde{R}},3} &=& -i (\widetilde{\alpha}^\dagger_{\mathbf{\widetilde{R}},3} \widetilde{\beta}_{\mathbf{\widetilde{R}},3} - \widetilde{\beta}^\dagger_{\mathbf{\widetilde{R}},3} \widetilde{\alpha}_{\mathbf{\widetilde{R}},3}) ~, 
\label{spinSW3SN}
\end{eqnarray}
with constraints $(\widetilde{\alpha}_{\mathbf{\widetilde{R}},3})^2=(\widetilde{\beta}_{\mathbf{\widetilde{R}},3})^2=\widetilde{\alpha}_{\mathbf{\widetilde{R}},3}\widetilde{\beta}_{\mathbf{\widetilde{R}},3}=\widetilde{\beta}_{\mathbf{\widetilde{R}},3}\widetilde{\alpha}_{\mathbf{\widetilde{R}},3}=\widetilde{\alpha}_{\mathbf{\widetilde{R}},3} \widetilde{\beta}_{\mathbf{\widetilde{R}},3}^\dagger=\widetilde{\beta}_{\mathbf{\widetilde{R}},3} \widetilde{\alpha}_{\mathbf{\widetilde{R}},3}^\dagger=0$ and $\widetilde{\beta}_{\mathbf{\widetilde{R}},3} \widetilde{\beta}_{\mathbf{\widetilde{R}},3}^\dagger=\widetilde{\alpha}_{\mathbf{\widetilde{R}},3}\widetilde{\alpha}_{\mathbf{\widetilde{R}},3}^\dagger=1-\widetilde{\alpha}_{\mathbf{\widetilde{R}},3}^\dagger \widetilde{\alpha}_{\mathbf{\widetilde{R}},3} -\widetilde{\beta}_{\mathbf{\widetilde{R}},3}^\dagger\widetilde{\beta}_{\mathbf{\widetilde{R}},3}$ for faithful representation of the Hilbert space. Likewise, the spin operators can be written for the other two sublattices.\cite{doi:10.1143/JPSJ.75.083701} Eq. \ref{Heff}, up to harmonic  orders in the bosons is,
\begin{equation}
        H_{sp} = 3 K_0 N +  H(\widetilde{\beta}_1,\widetilde{\alpha}_2) + H(\widetilde{\beta}_2,\widetilde{\alpha}_3) + H(\widetilde{\beta}_3,\widetilde{\alpha}_1) ~,
\end{equation}
with the following form of the Hamiltonian for each pair featuring above: 
\begin{eqnarray}
 H(\widetilde{\beta}_{\widetilde{\lambda}},\widetilde{\alpha}_{\widetilde{\rho}}) &=& 3 \sum_{\mathbf{k}} \big[\widetilde{\gamma}_{\mathbf{k}} ( J_0 \widetilde{\beta}^\dagger_{\mathbf{k},\widetilde{\lambda}} \widetilde{\alpha}^\dagger_{-\mathbf{k},\widetilde{\rho}} + (J_0-K_0) \widetilde{\beta}^\dagger_{\mathbf{k},\widetilde{\lambda}} \widetilde{\alpha}_{\mathbf{k},\widetilde{\rho}} )\nonumber\\
 &&~~~~~~~~+\text{h.c}\big]+3K_0 \sum_{\mathbf{k}} (\Bar{n}_{\widetilde{\beta},\mathbf{k},\widetilde{\lambda}}+\Bar{n}_{\widetilde{\alpha},\mathbf{k},\widetilde{\rho}}) \nonumber \\
\end{eqnarray}
where $\widetilde{\gamma}_{\mathbf{k}}$ is as defined below Eq. \ref{dispersion3SN} and $\Bar{n}_{\widetilde{\beta},\mathbf{k},\widetilde{\lambda}}=\widetilde{\beta}^\dagger_{\mathbf{k},\widetilde{\lambda}} \widetilde{\beta}_{\mathbf{k},\widetilde{\lambda}}$, $\Bar{n}_{\widetilde{\alpha},\mathbf{k},\widetilde{\rho}}=\widetilde{\alpha}^\dagger_{\mathbf{k},\widetilde{\rho}} \widetilde{\alpha}_{\mathbf{k},\widetilde{\rho}}$.
The above Hamiltonian can be diagonalized using Bogoliubov transformation,\cite{10.1093/ptep/ptaa151}
\begin{equation}
    \left(\begin{matrix} \widetilde{\alpha}_{\mathbf{k},\widetilde{\rho}}\\\widetilde{\beta}_{\mathbf{k},\widetilde{\lambda}}\\\widetilde{\alpha}_{-\mathbf{k},\widetilde{\rho}}^{\dagger}\\\widetilde{\beta}_{-\mathbf{k},\widetilde{\lambda}}^{\dagger} \end{matrix}\right) = T_{\mathbf{k}} \left(\begin{matrix} \widetilde{d}_{+, \mathbf{k}, \widetilde{\lambda} \widetilde{\rho}}\\\widetilde{d}_{-, \mathbf{k}, \widetilde{\lambda} \widetilde{\rho}}\\\widetilde{d}_{+, -\mathbf{k}, \widetilde{\lambda} \widetilde{\rho}}^{~\dagger}\\\widetilde{d}_{-, -\mathbf{k}, \widetilde{\lambda} \widetilde{\rho}}^{~\dagger} \end{matrix}\right) \text{ with } T_{\mathbf{k}}=\left(\begin{matrix} \widetilde{T}_{1,{\mathbf{k}}} & \widetilde{T}_{2,{\mathbf{k}}} \\ \widetilde{T}_{2,{\mathbf{k}}} & \widetilde{T}_{1,{\mathbf{k}}}\end{matrix}\right)
\label{Tdefn}
\end{equation}
where
\begin{eqnarray}
 \widetilde{T}_{1,{\mathbf{k}}}&=&\left(\begin{matrix} \frac{\widetilde{t}_{+,\mathbf{k}}}{\widetilde{\gamma}_{\mathbf{k}} \widetilde{N}_{+,\mathbf{k}}} & \frac{\widetilde{t}_{-,\mathbf{k}}}{\widetilde{\gamma}_{\mathbf{k}} \widetilde{N}_{-,\mathbf{k}}} \\ -\frac{\widetilde{t}_{+,\mathbf{k}}}{|\widetilde{\gamma}_{\mathbf{k}}| \widetilde{N}_{+,\mathbf{k}}} & \frac{\widetilde{t}_{-,\mathbf{k}}}{|\widetilde{\gamma}_{\mathbf{k}}| \widetilde{N}_{-,\mathbf{k}}}\end{matrix}\right) \nonumber\\
 \widetilde{T}_{2,{\mathbf{k}}}&=&\left(\begin{matrix} -\frac{\widetilde{\gamma}^{*}_{\mathbf{k}}J_0}{|\widetilde{\gamma}_{\mathbf{k}}| \widetilde{N}_{+,\mathbf{k}}} & \frac{\widetilde{\gamma}^{*}_{\mathbf{k}}J_0}{|\widetilde{\gamma}_{\mathbf{k}}| \widetilde{N}_{-,\mathbf{k}}} \\ \frac{J_0}{ \widetilde{N}_{+,\mathbf{k}}} & \frac{J_0}{ \widetilde{N}_{-,\mathbf{k}}}\end{matrix}\right) 
\end{eqnarray}
with $\widetilde{t}_{\pm,\mathbf{k}} = \pm (J_0-K_0) |\widetilde{\gamma}_{\mathbf{k}}|-\left(K_0+\frac{\widetilde{\omega}^s_{\pm,\mathbf{k}}}{3}\right)$, $\widetilde{N}_{\pm,\mathbf{k}} =\frac{2K_0}{|\widetilde{\gamma}_{\mathbf{k}}|} 
 \sqrt{1+{\omega}^{\prime}_{\pm,\mathbf{k}}\pm (1-J^{\prime})|\widetilde{\gamma}_{\mathbf{k}}|(2+{\omega}^{\prime}_{\pm,\mathbf{k}})+(1-2J^\prime)|\widetilde{\gamma}_{\mathbf{k}}|^2}$ where ${\omega}^{\prime}_{\pm,\mathbf{k}}=\frac{\widetilde{\omega}^s_{\pm,\mathbf{k}}}{3K_0}$, $J^\prime=\frac{J_0}{K_0}$ and the Hamiltonian and dispersions are given in Eq. \ref{Hsp3SN} and Eq. \ref{dispersion3SN}.
 
\section{\label{sec:AppendixTinSN} Temperature dependence of fractional change in sound speed in the spin nematic phases}

\subsection{\label{sec:AppendixTinFN} Ferronematic phase}
The coupling Hamiltonian for the ferro spin-nematic waves and the phonons are given by  Eq. \ref{H12} where the scattering vertices, $\mathcal{M}_{\mathbf{k},\mathbf{q_2}}^{(1)}$, $\mathcal{M}_{\mathbf{k},\mathbf{k}^{\prime},\mathbf{q_2}}^{(2)}$ and $\mathcal{M}_{\mathbf{k}}^{(2),0}$ are given by
\begin{align}
   \mathcal{M}_{\mathbf{k},\mathbf{q_2}}^{(1)}=\frac{1}{2}\sum_{\boldsymbol{\delta}}\frac{(1-e^{i\mathbf{k}\cdot{\boldsymbol{\delta}}})}{\sqrt{2 M N \omega_{0,\mathbf{k}}^{ph}}}~ \overline{\mathcal{M}}_{\mathbf{k},\mathbf{q_2}}^{(1)}(\boldsymbol{\delta})
   \label{M1} 
\end{align}
with
\begin{widetext}

\begin{align}
&\overline{\mathcal{M}}_{\mathbf{k},\mathbf{q_2}}^{(1)}(\boldsymbol{\delta})=\mathcal{R}_{\mathbf{k}+\mathbf{q_2}}\left[(e^{i\mathbf{q}_2\cdot{\boldsymbol{\delta}}}+e^{-i(\mathbf{k}+\mathbf{q}_2)\cdot{\boldsymbol{\delta}}})\left[-\frac{\partial J_{\boldsymbol{\delta}}}{\partial\boldsymbol{\delta}}\cdot\mathbf{e}_{\mathbf{k}}\left[\begin{matrix} 1 & 1 \\ 1 & 1\end{matrix}\right]+\frac{\partial K_{\boldsymbol{\delta}}}{\partial \boldsymbol{\delta}}\cdot\mathbf{e}_{\mathbf{k}}\left[\begin{matrix} 0 & 1 \\ 1 & 0\end{matrix}\right]\right]+(1+e^{-i\mathbf{k}\cdot{\boldsymbol{\delta}}})\frac{\partial K_{\boldsymbol{\delta}}}{\partial\boldsymbol{\delta}}\cdot\mathbf{e}_{\mathbf{k}}\left[\begin{matrix} 1 & 0 \\ 0 & 1\end{matrix}\right]\right]\mathcal{R}_{\mathbf{q_2}}
\label{M1bar}
\end{align}
and
\begin{eqnarray}
\mathcal{M}_{\mathbf{k},\mathbf{k}^{\prime},\mathbf{q_2}}^{(2)} &=& \sum_{\boldsymbol{\delta}} \frac{(1-e^{i\mathbf{k}\cdot{\boldsymbol{\delta}}})(1-e^{-i\mathbf{k}^{\prime}\cdot{\boldsymbol{\delta}}})}{8 M N \sqrt{\omega_{0,\mathbf{k}}^{ph}  \omega_{0,-\mathbf{k}^{\prime}}^{ph}}}\mathcal{R}_{\mathbf{k}-\mathbf{k}^{\prime}+\mathbf{q_2}}\bigg[\left(\mathbf{e}_{-\mathbf{k}^{\prime}}\cdot\frac{\partial^2 J_{\boldsymbol{\delta}}}{\partial{\boldsymbol{\delta}^{2}}}\cdot\mathbf{e}_{\mathbf{k}}\right)(e^{i\mathbf{q_2}\cdot\boldsymbol{\delta}}+e^{-i(\mathbf{k}-\mathbf{k}^{\prime}+\mathbf{q_2})\cdot\boldsymbol{\delta}})\left(\begin{matrix} 1 & 1 \\ 1 & 1\end{matrix}\right)\nonumber\\
&&~~~~~-\left(\mathbf{e}_{-\mathbf{k}^{\prime}}\cdot\frac{\partial^2 K_{\boldsymbol{\delta}}}{\partial{\boldsymbol{\delta}^{2}}}\cdot\mathbf{e}_{\mathbf{k}}\right)\bigg((e^{i\mathbf{q}_2\cdot{\boldsymbol{\delta}}}+e^{-i(\mathbf{k}-\mathbf{k}^{\prime}+\mathbf{q}_2)\cdot{\boldsymbol{\delta}}})\left(\begin{matrix} 0 & 1 \\ 1 & 0\end{matrix}\right)+(1+e^{-i(\mathbf{k}-\mathbf{k}^{\prime})\cdot{\boldsymbol{\delta}}})\left(\begin{matrix} 1 & 0 \\ 0 & 1\end{matrix}\right)\bigg)\bigg]\mathcal{R}_{\mathbf{q}_2}
\end{eqnarray}
\begin{eqnarray}
\mathcal{M}_{\mathbf{k}}^{(2),0} &=& \sum_{\boldsymbol{\delta}}\frac{1}{2 M  \omega_{0,\mathbf{k}}^{ph}}(1-e^{i\mathbf{k}\cdot{\boldsymbol{\delta}}})(1-e^{-i\mathbf{k}\cdot{\boldsymbol{\delta}}}) \left(\mathbf{e}_{-\mathbf{k}}\cdot\frac{\partial^2 K_{\boldsymbol{\delta}}}{\partial{\boldsymbol{\delta}^{2}}}\cdot\mathbf{e}_{\mathbf{k}}\right).
\label{eq:wideeq}
\end{eqnarray}
The matrix $\mathcal{R}_{\mathbf{k}}$ is  defined in Eq. \ref{Rdefn}. $J_{\boldsymbol{\delta}}$ and $K_{\boldsymbol{\delta}}$ denote the coupling constants depend on the relative separation ($\boldsymbol{\delta}$) between the two nearest neighbour spins. The resultant temperature dependence of the fractional change in sound speed is given by Eq. \ref{FNdvbyv} with
\begin{eqnarray}
\Bar{c}_1 =&& \sum_{\boldsymbol{\delta}} \frac{1}{M v_0^2} (\hat{\mathbf{q}}\cdot\boldsymbol{\delta})^2  \left(\mathbf{e}_{-\mathbf{q}}\cdot\frac{\partial^2 K_{\boldsymbol{\delta}}}{\partial{\boldsymbol{\delta}^{2}}}\cdot\mathbf{e}_{\mathbf{q}}\right)-\frac{V}{32 \pi^2 M N v_0^2} \int_{\text{BZ}} dq_2 d\theta_{\hat{\mathbf{q}}_2} q_2 \sum_{\boldsymbol{\delta},\boldsymbol{\delta}^{\prime}} \frac{(\hat{\mathbf{q}}\cdot\boldsymbol{\delta})(\hat{\mathbf{q}}\cdot\boldsymbol{\delta}^{\prime})}{\omega^s_{\mathbf{q}_2}} \overline{\mathcal{M}}_{0,\mathbf{q}_2}^{(1),12}(\boldsymbol{\delta}) \overline{\mathcal{M}}_{0,\mathbf{q}_2}^{(1),21}(\boldsymbol{\delta}^{\prime})\nonumber\\
&&+\frac{V}{8 \pi^2 M N v_0^2} \int_{\text{BZ}} dq_2 d\theta_{\hat{\mathbf{q}}_2}q_2\sum_{\boldsymbol{\delta}} (\hat{\mathbf{q}}\cdot\boldsymbol{\delta})^2 \bigg[\left(\mathbf{e}_{-\mathbf{q}}\cdot\frac{\partial^2 J_{\boldsymbol{\delta}}}{\partial \boldsymbol{\delta}^2}\cdot\mathbf{e}_{\mathbf{q}}\right)\text{cos}(\mathbf{q}_2\cdot\boldsymbol{\delta})(\text{cosh}2\Bar{\theta}_{\mathbf{q}_2}+\text{sinh}2\Bar{\theta}_{\mathbf{q}_2})\nonumber\\
&&-\left(\mathbf{e}_{-\mathbf{q}}\cdot\frac{\partial^2 K_{\boldsymbol{\delta}}}{\partial \boldsymbol{\delta}^2}\cdot\mathbf{e}_{\mathbf{q}}\right)(\text{cosh}2\Bar{\theta}_{\mathbf{q}_2}+\text{cos}(\mathbf{q}_2\cdot\boldsymbol{\delta})\text{sinh}2\Bar{\theta}_{\mathbf{q}_2})\bigg]\nonumber\\
\Bar{c}_2 =&& \frac{V k_B^3}{32 \pi^2 M N v_0^2 \Bar{c}_s^4} \int_0^{\infty}  dx \int_0^{2\pi} d\theta_{\hat{\mathbf{q}}_2} \bigg[ \frac{2 x^3 e^{x} \Bar{\mathcal{A}}_{+}(\hat{\mathbf{q}},\hat{\mathbf{q}_2}) \Bar{c}_s^2 \cos^2{\theta_{\mathbf{q} \mathbf{q}_2}}}{(e^{x}-1)^2(v_0^2-\Bar{c}_s^2 \cos^2{\theta_{\mathbf{q} \mathbf{q}_2}})}
+ \frac{2 x^2}{(e^x-1)} \bigg(-\Bar{\mathcal{A}}_{-}(\hat{\mathbf{q}},\hat{\mathbf{q}}_2)+2 \Bar{c}_s \Bar{\mathcal{C}}(\hat{\mathbf{q}},\hat{\mathbf{q}}_2)\bigg) \bigg]\nonumber
\end{eqnarray}
where $q_2=|\mathbf{q}_2|$, $V$ is the total volume of the lattice, $\theta_{\mathbf{q} \mathbf{q}_2}$ is the angle between $\hat{\mathbf{q}}$ and $\hat{\mathbf{q}}_2$, the integration limits in $\Bar{c}_2$ can be extended upto infinity due to exponential damping factor and 
\begin{eqnarray*}
\Bar{\mathcal{A}}_{\pm}(\hat{\mathbf{q}},\hat{\mathbf{q}_2}) =&& \sum_{\boldsymbol{\delta},\boldsymbol{\delta}^{\prime}} (\hat{\mathbf{q}}\cdot\boldsymbol{\delta})(\hat{\mathbf{q}}\cdot\boldsymbol{\delta}^{\prime})\bigg[\left(\frac{\partial J_{\boldsymbol{\delta}}}{\partial \boldsymbol{\delta}}\cdot\mathbf{e}_{-\mathbf{q}}\right)\frac{3 K_0}{\Bar{c}_s}-\left(\frac{\partial K_{\boldsymbol{\delta}}}{\partial \boldsymbol{\delta}}\cdot\mathbf{e}_{-\mathbf{q}}\right)\left(\frac{3 K_0}{\Bar{c}_s} \pm \frac{(\hat{\mathbf{q}}_2\cdot\boldsymbol{\delta})^2}{\Bar{c}_s} 6 (K_0-J_0)\right)\bigg]\nonumber\\
&&~~~~\times\bigg[\left(\frac{\partial J_{\boldsymbol{\delta}^{\prime}}}{\partial \boldsymbol{\delta}^{\prime}}\cdot\mathbf{e}_{\mathbf{q}}\right)\frac{3 K_0}{\Bar{c}_s}-\left(\frac{\partial K_{\boldsymbol{\delta}^{\prime}}}{\partial \boldsymbol{\delta}^{\prime}}\cdot\mathbf{e}_{\mathbf{q}}\right)\left(\frac{3 K_0}{\Bar{c}_s}\pm\frac{(\hat{\mathbf{q}}_2\cdot\boldsymbol{\delta^{\prime}})^2}{\Bar{c}_s} 6 (K_0-J_0)\right)\bigg]\\
\Bar{\mathcal{C}}(\hat{\mathbf{q}},\hat{\mathbf{q}_2}) =&& \sum_{\boldsymbol{\delta}} (\hat{\mathbf{q}}\cdot\boldsymbol{\delta})^2 \bigg[-\left(\mathbf{e}_{-\mathbf{q}}\cdot\frac{\partial^2 J_{\boldsymbol{\delta}}}{\partial \boldsymbol{\delta}^2}\cdot\mathbf{e}_{\mathbf{q}}\right)\frac{3 K_0}{\Bar{c}_s}+\left(\mathbf{e}_{-\mathbf{q}}\cdot\frac{\partial^2 K_{\boldsymbol{\delta}}}{\partial \boldsymbol{\delta}^2}\cdot\mathbf{e}_{\mathbf{q}}\right)\left(\frac{3 K_0}{\Bar{c}_s}+\frac{(\hat{\mathbf{q}}_2\cdot\boldsymbol{\delta})^2}{\Bar{c}_s} 6 (K_0-J_0)\right)\bigg]\nonumber\\
\end{eqnarray*}
\end{widetext}

To obtain the sound attenuation $\propto \Sigma(\mathbf{q},\omega)/v_0$, we note that at the leading order only $\Sigma_1$ has an imaginary part. From Eq. \ref{M1}, it is fairly straight forward to show that it is proportional to $\omega_{0,\mathbf{q}}^{ph}$ and hence disappears for the acoustic phonons in the limit of ${\bf q}\rightarrow 0$.\cite{cottam1974spin}

\subsection{\label{sec:AppendixTin3SN} Three sublattice nematic phase}

For the three sublattice nematic, the spin-phonon coupling Hamiltonian in Eq. \ref{HTot3SN} is given by
\begin{align}
   H_{sp-ph,3SN} =H_{1,3SN} + H_{2,3SN} 
   \label{eq_Hspph-3SN}
\end{align}
where the two terms are similar to Eqs. \ref{H12} are given by 
\begin{align}
H_{1,3SN} =& \sum_{\mathbf{k},\mathbf{q_2},\widetilde{\lambda} \widetilde{\rho}} \widetilde{\psi}^{\dagger}_{\mathbf{k}+\mathbf{q_2},\widetilde{\lambda} \widetilde{\rho}} \widetilde{\mathcal{M}}_{\mathbf{k}+\mathbf{q_2},\mathbf{q_2}}^{(1)} \widetilde{\psi}_{\mathbf{q_2},\widetilde{\lambda} \widetilde{\rho}} A_{\mathbf{k}}
\end{align}

\begin{align}
H_{2,3SN} =& \sum_{\mathbf{k},\mathbf{k}^{\prime},\mathbf{q_2},\widetilde{\lambda} \widetilde{\rho}} \widetilde{\psi}^{\dagger}_{\mathbf{k}-\mathbf{k}^{\prime}+\mathbf{q_2},\widetilde{\lambda} \widetilde{\rho}} \widetilde{\mathcal{M}}_{\mathbf{k},\mathbf{k}^{\prime},\mathbf{q_2}}^{(2)} \widetilde{\psi}_{\mathbf{q_2},\widetilde{\lambda} \widetilde{\rho}} A_{\mathbf{k}} A_{-\mathbf{k}^{\prime}}
\nonumber\\
&+\sum_{\mathbf{k}} \widetilde{\mathcal{M}}_{\mathbf{k}}^{(2),0} A_{\mathbf{k}} A_{-\mathbf{k}} ~,
\label{H123SN}
\end{align}
where $\widetilde{\psi}_{\mathbf{k},\widetilde{\lambda} \widetilde{\rho}}=\bigg(\widetilde{d}_{+,\mathbf{k},\widetilde{\lambda} \widetilde{\rho}} , \text{  }\widetilde{d}_{-,\mathbf{k},\widetilde{\lambda} \widetilde{\rho}} ,\text{  } \widetilde{d}^{~\dagger}_{+,-\mathbf{k},\widetilde{\lambda}\widetilde{\rho}} ,\text{  } \widetilde{d}^{~\dagger}_{-,-\mathbf{k},\widetilde{\lambda} \widetilde{\rho}}\bigg)^T$ and 
\begin{align}
\widetilde{\mathcal{M}}_{\mathbf{k}+\mathbf{q_2},\mathbf{q_2}}^{(1)}=&\sum_{\boldsymbol{\widetilde{\delta}}}\sqrt{\frac{1}{2 M N \omega_{0,\mathbf{k}}^{ph}}} (1-e^{i\mathbf{k}\cdot{\boldsymbol{\widetilde{\delta}}}}) \overline{\widetilde{\mathcal{M}}}_{\mathbf{k}+\mathbf{q_2},\mathbf{q_2}}^{(1)}(\boldsymbol{\widetilde{\delta}})\label{3SNM1}
\end{align}
where
\begin{widetext}
\begin{align}
\overline{\widetilde{\mathcal{M}}}_{\mathbf{k}+\mathbf{q_2},\mathbf{q_2}}^{(1)}(\boldsymbol{\widetilde{\delta}}) =& T^\dagger_{\mathbf{k}+\mathbf{q_2}}\Bigg[ \left(\frac{\partial J_{\boldsymbol{\widetilde{\delta}}}}{\partial (-\boldsymbol{\widetilde{\delta}})}\cdot\mathbf{e}_{\mathbf{k}}\right) e^{i\mathbf{q}_2\cdot{\boldsymbol{\widetilde{\delta}}}}\left(\begin{matrix} 0 & 0 & 0 & 0 \\ 1 & 0 & 1 & 0 \\0 & 0 & 0 & 0 \\ 1 & 0 & 1 & 0\end{matrix}\right)+\left(\frac{\partial K_{\boldsymbol{\widetilde{\delta}}}}{\partial (-\boldsymbol{\widetilde{\delta}})}\cdot\mathbf{e}_{\mathbf{k}}\right)\left(\begin{matrix} e^{-i\mathbf{k}\cdot{\boldsymbol{\widetilde{\delta}}}} & -e^{-i(\mathbf{k}+\mathbf{q}_2)\cdot{\boldsymbol{\widetilde{\delta}}}} & 0 & 0\\ -e^{i \mathbf{q}_2\cdot{\boldsymbol{\widetilde{\delta}}}} & 1 & 0 & 0\\0 & 0 & 0 & 0\\0 & 0 & 0 & 0\end{matrix}\right)\Bigg]T_{\mathbf{q}_2}\label{3SNM1bar}
\end{align}
and
\begin{align}
\widetilde{\mathcal{M}}_{\mathbf{k},\mathbf{k}^{\prime},\mathbf{q_2}}^{(2)} =& \sum_{\boldsymbol{\widetilde{\delta}}} \frac{1}{4 M N \sqrt{\omega_{0,\mathbf{k}}^{ph}  \omega_{0,-\mathbf{k}^{\prime}}^{ph}}}(1-e^{i\mathbf{k}\cdot{\boldsymbol{\widetilde{\delta}}}})(1-e^{-i\mathbf{k}^{\prime}\cdot{\boldsymbol{\widetilde{\delta}}}}) \overline{\widetilde{\mathcal{M}}}^{(2)}_{\mathbf{k},\mathbf{k}^\prime,\mathbf{q_2}}(\boldsymbol{\widetilde{\delta}})
\end{align}
where
\begin{align}
\overline{\widetilde{\mathcal{M}}}^{(2)}_{\mathbf{k},\mathbf{k}^\prime,\mathbf{q_2}}(\boldsymbol{\widetilde{\delta}})=&T^{\dagger}_{\mathbf{k}-\mathbf{k}^{\prime}+\mathbf{q}_2} \Bigg[ \mathbf{e}_{-\mathbf{k}^{\prime}}\cdot\frac{\partial^2 J_{\boldsymbol{\widetilde{\delta}}}}{\partial{\boldsymbol{\widetilde{\delta}}^{2}}}\cdot\mathbf{e}_{\mathbf{k}} e^{i\mathbf{q}_2\cdot{\boldsymbol{\widetilde{\delta}}}}\left[\begin{matrix} 0 & 0 & 0 & 0 \\ 1 & 0 & 1 & 0 \\0 & 0 & 0 & 0 \\ 1 & 0 & 1 & 0\end{matrix}\right]+\mathbf{e}_{-\mathbf{k}^{\prime}}\cdot\frac{\partial^2 K_{\boldsymbol{\widetilde{\delta}}}}{\partial{\boldsymbol{\widetilde{\delta}}^{2}}}\cdot\mathbf{e}_{\mathbf{k}}\left[\begin{matrix} e^{-i(\mathbf{k}-\mathbf{k}^{\prime})\cdot{\boldsymbol{\widetilde{\delta}}}} & -e^{-i(\mathbf{k}-\mathbf{k}^{\prime}+\mathbf{q}_2)\cdot{\boldsymbol{\widetilde{\delta}}}} & 0 & 0\\ -e^{i \mathbf{q}_2\cdot{\boldsymbol{\widetilde{\delta}}}} & 1 & 0 & 0\\0 & 0 & 0 & 0\\0 & 0 & 0 & 0\end{matrix}\right]\Bigg]T_{\mathbf{q}_2}
\end{align}

\begin{align}
\widetilde{\mathcal{M}}_{\mathbf{k}}^{(2),0} =& \sum_{\boldsymbol{\widetilde{\delta}}}\frac{1}{4 M  \omega_{0,\mathbf{k}}^{ph}}(1-e^{i\mathbf{k}\cdot{\boldsymbol{\widetilde{\delta}}}})(1-e^{-i\mathbf{k}\cdot{\boldsymbol{\widetilde{\delta}}}}) \left(\mathbf{e}_{-\mathbf{k}}\cdot\frac{\partial^2 K_{\boldsymbol{\widetilde{\delta}}}}{\partial{\boldsymbol{\widetilde{\delta}}^{2}}}\cdot\mathbf{e}_{\mathbf{k}}\right)
\label{3SNM2}
\end{align}
\end{widetext}
Here, the matrix $T_{\mathbf{k}}$ is as defined in Eq. \ref{Tdefn} and $\boldsymbol{\widetilde{\delta}}$ are the displacements mentioned below Eq. \ref{dispersion3SN}.

The phonon self energy is then calculated perturbatively (in spin-phonon interaction) and the leading contributions are given by:
\begin{align}
    \Sigma(\mathbf{q}, i\Omega_n)&= \sum_{\mathbf{k}} \widetilde{\Sigma}_{\mathbf{k}}(\mathbf{q}, i\Omega_n)\nonumber\\
    &=\sum_{\mathbf{k}}[\widetilde{\Sigma}_{1 \mathbf{k}}(\mathbf{q}, i\Omega_n)+\widetilde{\Sigma}_{2 \mathbf{k}}(\mathbf{q}, i\Omega_n)] ~,
\end{align}
where
\begin{align}
&\widetilde{\Sigma}_{1 \mathbf{k}} (\mathbf{q},i \Omega_{n}) \approx  \frac{3}{2} \sum_{\widetilde{\mu},\widetilde{\nu}\in\{+,-\}} \nonumber\\ &\bigg[-\frac{1}{\beta} \mathcal{N}_{2,\mathbf{q},\mathbf{k}}^{\widetilde{\mu} \widetilde{\nu}} \sum_{\omega_1} \widetilde{G}^{m,\widetilde{\mu}}_{0}(\mathbf{q}+\mathbf{k},i \Omega_{n}+i \omega_1) \widetilde{G}^{m,\widetilde{\nu}}_{0}(\mathbf{k},i \omega_1) 
\nonumber\\
&-\frac{1}{\beta} \mathcal{N}_{3,\mathbf{q},\mathbf{k}}^{\widetilde{\mu}\widetilde{\nu}} \sum_{\omega_1} \widetilde{G}^{m,\widetilde{\mu}}_{0}(-\mathbf{q}+\mathbf{k},-i \Omega_{n}-i \omega_1) \widetilde{G}^{m,\widetilde{\nu}}_{0}(-\mathbf{k},i \omega_1)\nonumber\\ 
&-\frac{1}{\beta} \mathcal{N}_{1,\mathbf{q},\mathbf{k}}^{\widetilde{\mu} \widetilde{\nu}} \sum_{\omega_1} \widetilde{G}^{m,\widetilde{\mu}}_{0}(\mathbf{q}+\mathbf{k},i \Omega_{n}-i \omega_1) \widetilde{G}^{m,\widetilde{\nu}}_{0}(-\mathbf{k},i \omega_1)  \bigg]\label{MatsubaraSelfEnergy3SNH1}\\
&\widetilde{\Sigma}_{2 \mathbf{k}} (\mathbf{q},i \Omega_{n}) \approx  2 \widetilde{\mathcal{M}}_{\mathbf{q}}^{(2),0}+6  \sum_{\tilde{i}=\{1 \text{ to } 4\}} \widetilde{\mathcal{M}}_{\mathbf{q},\mathbf{q},\mathbf{k}}^{(2),\tilde{i}\tilde{i}} \langle \widetilde{\psi}^{\dagger \tilde{i}}_{\mathbf{k}}\widetilde{\psi}^{~\tilde{i}}_{\mathbf{k}}\rangle  ~,
\label{MatsubaraSelfEnergy3SNH2}
\end{align}
where the $\widetilde{\lambda}\widetilde{\rho}$ indices are suppressed since each pair gives the same contribution and has been accounted by appropriate multiplicative factors in the expressions above, $\widetilde{G}^{m,+}_{0}(\mathbf{q},i \Omega_{n})$ and $\widetilde{G}^{m,-}_{0}(\mathbf{q},i \Omega_{n})$ are the bare propagators for the $\widetilde{d}_{+,\mathbf{q}}$ and $\widetilde{d}_{-,\mathbf{q}}$ bosons: $\widetilde{G}^{m,+}_{0}(\mathbf{q},i \Omega_{n})=\frac{1}{i \Omega_{n}-\widetilde{\omega}^s_{+,\mathbf{q}}}$ and $\widetilde{G}^{m,-}_{0}(\mathbf{q},i \Omega_{n})=\frac{1}{i \Omega_{n}-\widetilde{\omega}^s_{-,\mathbf{q}}}$ and  $\mathcal{N}$ in $\Sigma_1$ stands for the appropriate matrix elements of $\mathcal{O}((\widetilde{\mathcal{M}}_{\mathbf{k},\mathbf{q_2}}^{(1)})^2)$. The sum over Matsubara frequencies can be performed using the techniques in Ref. \onlinecite{bruus2004many}. Converting the momentum sums into integrals one obtains  Eq. \ref{eq_3sndvv2} in the main text.

{For $J_0/K_0\ll 1$ and temperatures much less than the gap scale of $\widetilde{\omega}^s_{+,\mathbf{q}}$, the temperature dependent contribution to Eq. \ref{MatsubaraSelfEnergy3SNH1} due to both the internal lines in Fig. \ref{diagrams} belonging to the $\widetilde{\omega}^s_{+,\mathbf{q}}$ branch, the contribution is exponentially damped due to the finite gap. When one branch is $\widetilde{\omega}^s_{+,\mathbf{q}}$ and the other is $\widetilde{\omega}^s_{-,\mathbf{q}}$, the temperature dependence obtained is similar to that obtained for both the branches being $\widetilde{\omega}^s_{-,\mathbf{q}}$ albeit with different prefactors.  Likewise, the contribution to the second term of Eq. \ref{MatsubaraSelfEnergy3SNH2} due to the $\widetilde{\omega}^s_{+,\mathbf{q}}$ would be exponentially damped.}

We now list the detailed expressions obtained by considering only the $\widetilde{\omega}^s_{-,\mathbf{q}}$ branch for obtaining the self energy. Now, to scale out the temperature dependence due to the Bose distribution function, we perform the standard change of variables $x=\beta \widetilde{\omega}^s_{-,\mathbf{k}}$ (where $\beta=1/(k_B T)$) and obtain the temperature dependence. This changes the upper limit of the first term in Eq. \ref{eq_3sndvv2} and the lower limit of the second term in Eq. \ref{eq_3sndvv2} to $\beta c_s k^*=\beta \widetilde{c_s} (k^*)^2= 6 \beta J_0$ for integration over the variable $x$.  In the limit where $\beta J_0 \gg 1$, only the contribution from the first integral in Eq. \ref{eq_3sndvv2} with the linear dispersion would be important, giving the following expressions for the pre-factors in Eq. \ref{eq_3sndvv3}, 
\begin{widetext}
\begin{align}
\widetilde{c}_1 =& \sum_{\boldsymbol{\widetilde{\delta}}} \frac{1}{2 M v_0^2} (\hat{\mathbf{q}}\cdot\boldsymbol{\widetilde{\delta}})^2  \left(\mathbf{e}_{-\mathbf{q}}\cdot\frac{\partial^2 K_{\boldsymbol{\widetilde{\delta}}}}{\partial{\boldsymbol{\widetilde{\delta}}^{2}}}.\mathbf{e}_{\mathbf{q}}\right) - \frac{3V}{16\pi^2}\lim_{\mathbf{q} \xrightarrow{} 0}\int_{\text{BZ}}  dk d\theta_{\hat{\mathbf{k}}}k \frac{(\mathcal{N}_{1,\mathbf{q},\mathbf{k}}^{--}+\mathcal{N}_{3,\mathbf{q},\mathbf{k}}^{--})}{\widetilde{\omega}^s_{-,\mathbf{k}}\omega_{0,\mathbf{q}}^{ph}}\nonumber\\
&+\frac{3 V}{8\pi^2 M N v_0^2} \int_{\text{BZ}}  dk d\theta_{\hat{\mathbf{k}}}k \sum_{\boldsymbol{\widetilde{\delta}}} (\hat{\mathbf{q}}\cdot\boldsymbol{\widetilde{\delta}})^2 \overline{\widetilde{\mathcal{M}}}^{(2),44}_{0,0,\mathbf{k}}(\boldsymbol{\widetilde{\delta}})
\label{c1tilde}
\end{align}

\begin{align}
\widetilde{c}_2 =& \frac{3 V k_B^3}{4\pi^2 M N v_0^2 c_s^4} \int_{0}^{\infty}dx\int_{0}^{2\pi}d\theta_{\hat{\mathbf{k}}} \bigg[ \frac{x^3 e^{x}c_s \cos{\theta_{\mathbf{q} \mathbf{k}}}\mathcal{A}_{+}(\hat{\mathbf{q}},\hat{\mathbf{k}})K_0}{(e^{x}-1)^2(v_0-c_s \cos{\theta_{\mathbf{q} \mathbf{k}}})J_0} + \frac{x^2}{(e^x-1)} \big(-\mathcal{A}_{-}(\hat{\mathbf{q}},\hat{\mathbf{k}})\frac{K_0}{J_0}+ c_s \sqrt{\frac{K_0}{2J_0}}\mathcal{C}(\hat{\mathbf{q}},\hat{\mathbf{k}})\big) \bigg]
\end{align}
where $k=|\mathbf{k}|$, $\theta_{\mathbf{q} \mathbf{k}}$ is the angle between $\hat{\mathbf{q}}$ and $\hat{\mathbf{k}}$, the integration limits in $\widetilde{c}_2$ have been extended to infinity in the $\beta J_0\gg 1$ limit and
\begin{align}
\mathcal{A}_{\pm}(\hat{\mathbf{q}},\hat{\mathbf{k}}) =& \sum_{\boldsymbol{\widetilde{\delta}},\boldsymbol{\widetilde{\delta}}^{\prime}} (\hat{\mathbf{q}}\cdot\boldsymbol{\widetilde{\delta}})(\hat{\mathbf{q}}\cdot\boldsymbol{\widetilde{\delta}}^{\prime})\bigg[\frac{1}{4}\frac{\partial J_{\boldsymbol{\widetilde{\delta}}}}{\partial \boldsymbol{\widetilde{\delta}}}\cdot\mathbf{e}_{-\mathbf{q}}\pm\frac{\partial K_{\boldsymbol{\widetilde{\delta}}}}{\partial \boldsymbol{\widetilde{\delta}}}\cdot\mathbf{e}_{-\mathbf{q}}\left(\frac{J_0}{2K_0}\right)(\hat{\mathbf{k}}\cdot\boldsymbol{\widetilde{\delta}})^2 \bigg]\bigg[\frac{1}{4}\frac{\partial J_{\boldsymbol{\widetilde{\delta}}^{\prime}}}{\partial \boldsymbol{\widetilde{\delta}}^{\prime}}\cdot\mathbf{e}_{\mathbf{q}}\pm\frac{\partial K_{\boldsymbol{\widetilde{\delta}}^{\prime}}}{\partial \boldsymbol{\widetilde{\delta}}^{\prime}}\cdot\mathbf{e}_{\mathbf{q}}\left(\frac{J_0}{2K_0}\right)(\hat{\mathbf{k}}\cdot\boldsymbol{\widetilde{\delta}}^{\prime})^2 \bigg]
\end{align}

\begin{align}
\mathcal{C}(\hat{\mathbf{q}},\hat{\mathbf{k}}) =& \sum_{\boldsymbol{\widetilde{\delta}}} (\hat{\mathbf{q}}\cdot\boldsymbol{\widetilde{\delta}})^2\bigg[\frac{1}{4}\left(\mathbf{e}_{-\mathbf{q}}\cdot\frac{\partial^2 J_{\boldsymbol{\widetilde{\delta}}}}{\partial \boldsymbol{\widetilde{\delta}}^2}\cdot\mathbf{e}_{\mathbf{q}}\right)+\left(\mathbf{e}_{-\mathbf{q}}\cdot\frac{\partial^2 K_{\boldsymbol{\widetilde{\delta}}}}{\partial \boldsymbol{\widetilde{\delta}}^2}\cdot\mathbf{e}_{\mathbf{q}}\right)\left(\frac{J_0}{2K_0}\right)(\hat{\mathbf{k}}\cdot\boldsymbol{\widetilde{\delta}})^2\bigg]
\end{align}

 For $\beta J_0 \lesssim 1$, the leading temperature dependence is obtained from the second integral in Eq. \ref{eq_3sndvv2} with the quadratic dispersion, $\frac{\Delta v}{v} = \widetilde{c}_1+\mathcal{I}_1+\mathcal{I}_2+\mathcal{I}_3$
where $\widetilde{c}_1$ is given in Eq. \ref{c1tilde} and
\begin{align}
    \mathcal{I}_1 =& -\frac{3 V}{8\pi^2 M N v_0^2 \widetilde{c}_s}\int_{6\beta J_0}^{\infty}\frac{dx}{x(e^x-1)} \int d\theta_{\hat{\mathbf{k}}} \widetilde{\mathcal{A}}_{-}(\hat{\mathbf{q}},\hat{\mathbf{k}}) \nonumber\\
    \mathcal{I}_2 =& \frac{27 V}{4\pi^2 M N v_0^2}\int_{6\beta J_0}^{\infty}\frac{dx \sqrt{x} e^x} {\sqrt{\beta}(e^x-1)^2} \int d\theta_{\hat{\mathbf{k}}} \frac{\widetilde{\mathcal{A}}_{+}(\hat{\mathbf{q}},\hat{\mathbf{k}})\cos{\theta_{\mathbf{q} \mathbf{k}}}}{v_0\sqrt{\widetilde{c}_s}}\nonumber\\
    \mathcal{I}_3 =& \frac{9 V}{8\sqrt{2}\pi^2 M N v_0^2 \widetilde{c}_s}\int_{6\beta J_0}^{\infty}\frac{dx} {\beta(e^x-1)} \int d\theta_{\hat{\mathbf{k}}} \widetilde{\mathcal{C}}(\hat{\mathbf{q}},\hat{\mathbf{k}})
    \label{3SNintegrals}
\end{align}
where the upper limit can be taken to infinity due to exponential damping and  
\begin{align}
\widetilde{\mathcal{A}}_{+}(\hat{\mathbf{q}},\hat{\mathbf{k}}) =&  \sum_{\boldsymbol{\widetilde{\delta}},\boldsymbol{\widetilde{\delta}}^{\prime}} (\hat{\mathbf{q}}\cdot\boldsymbol{\widetilde{\delta}})(\hat{\mathbf{q}}\cdot\boldsymbol{\widetilde{\delta}}^{\prime})\bigg[\frac{1}{6}\frac{\partial J_{\boldsymbol{\widetilde{\delta}}}}{\partial \boldsymbol{\widetilde{\delta}}}\cdot\mathbf{e}_{-\mathbf{q}}+\frac{\partial K_{\boldsymbol{\widetilde{\delta}}}}{\partial \boldsymbol{\widetilde{\delta}}}\cdot\mathbf{e}_{-\mathbf{q}}\left(\frac{J_0}{K_0}\right)(\hat{\mathbf{k}}\cdot\boldsymbol{\widetilde{\delta}})^2 \bigg]\bigg[\frac{1}{6}\frac{\partial J_{\boldsymbol{\widetilde{\delta}}^{\prime}}}{\partial \boldsymbol{\widetilde{\delta}}^\prime}\cdot\mathbf{e}_{\mathbf{q}}+\frac{\partial K_{\boldsymbol{\widetilde{\delta}}^{\prime}}}{\partial \boldsymbol{\widetilde{\delta}}^\prime}\cdot\mathbf{e}_{\mathbf{q}}\left(\frac{J_0}{K_0}\right)(\hat{\mathbf{k}}\cdot\boldsymbol{\widetilde{\delta}}^{\prime})^2 \bigg] \\
\widetilde{\mathcal{A}}_{-}(\hat{\mathbf{q}},\hat{\mathbf{k}}) =&  \sum_{\boldsymbol{\widetilde{\delta}},\boldsymbol{\widetilde{\delta}}^{\prime}} (\hat{\mathbf{q}}\cdot\boldsymbol{\widetilde{\delta}})(\hat{\mathbf{q}}\cdot\boldsymbol{\widetilde{\delta}}^{\prime})\bigg[\frac{1}{2}\frac{\partial J_{\boldsymbol{\widetilde{\delta}}}}{\partial \boldsymbol{\widetilde{\delta}}}\cdot\mathbf{e}_{-\mathbf{q}}-\frac{\partial K_{\boldsymbol{\widetilde{\delta}}}}{\partial \boldsymbol{\widetilde{\delta}}}\cdot\mathbf{e}_{-\mathbf{q}}\left(\frac{J_0}{K_0}\right)(\hat{\mathbf{k}}\cdot\boldsymbol{\widetilde{\delta}})^2 \bigg]\bigg[\frac{1}{2}\frac{\partial J_{\boldsymbol{\widetilde{\delta}}^{\prime}}}{\partial \boldsymbol{\widetilde{\delta}}^\prime}\cdot\mathbf{e}_{\mathbf{q}}-\frac{\partial K_{\boldsymbol{\widetilde{\delta}}^{\prime}}}{\partial \boldsymbol{\widetilde{\delta}}^\prime}\cdot\mathbf{e}_{\mathbf{q}}\left(\frac{J_0}{K_0}\right)(\hat{\mathbf{k}}\cdot\boldsymbol{\widetilde{\delta}}^{\prime})^2 \bigg]
\end{align}

\begin{align}
\widetilde{\mathcal{C}}(\hat{\mathbf{q}},\hat{\mathbf{k}}) =& \sum_{\boldsymbol{\widetilde{\delta}}} (\hat{\mathbf{q}}\cdot\boldsymbol{\widetilde{\delta}})^2 \bigg[\frac{1}{6}\left(\mathbf{e}_{-\mathbf{q}}\cdot\frac{\partial^2 J_{\boldsymbol{\widetilde{\delta}}}}{\partial \boldsymbol{\widetilde{\delta}}^2}\cdot\mathbf{e}_{\mathbf{q}}\right)+\left(\mathbf{e}_{-\mathbf{q}}\cdot\frac{\partial^2 K_{\boldsymbol{\widetilde{\delta}}}}{\partial \boldsymbol{\widetilde{\delta}}^2}\cdot\mathbf{e}_{\mathbf{q}}\right)\left(\frac{J_0}{K_0}\right)(\hat{\mathbf{k}}\cdot\boldsymbol{\widetilde{\delta}})^2\bigg]
\end{align}
\end{widetext}
At exactly $J_0=0$, the lower limit of the integrals in Eq. \ref{3SNintegrals} is zero and the integrals diverge suggesting an instability at the $J_0=0$ point. At finite but small $J_0$, on performing numerical integration, we see the approximate temperature dependence as mentioned in Eq. \ref{eq_3sndvv3}.

\section{Details of the Landau-Ginzburg Theory}
\label{appen_lg}

\begin{figure}
\begin{center}
	\includegraphics[width=0.7\linewidth,keepaspectratio]{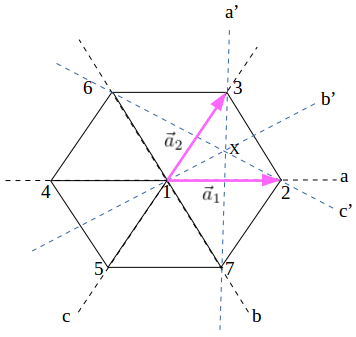}
\end{center}	
\caption{Triangular lattice of magnetic $\text{Ni}^{2+}$. Axes for the different symmetry transformations of $D_{3d}$ are shown}
\label{symmetryaxes} 
\end{figure}

The point group symmetry of ${\text{NiGa}_2\text{S}_4}$ is $\text{D}_{3d}$. This, along with translations by lattice vectors ($\mathbf{a}_1$ and $\mathbf{a}_2$ in Fig. \ref{symmetryaxes} in the plane of the triangular lattice of magnetic atoms) generate the set of relevant symmetry operations. Notably, the point group $\text{D}_{3d}$ is $\text{D}_3 \otimes C_i$ (where $C_i$ consists of identity($E$) and inversion($I$) operations).\cite{hamermesh1989group} $\text{D}_3$ consists of a $3$-fold rotation axis and a $2$-fold rotation axis perpendicular to the $3$-fold axis (The $2$-fold rotation axes $a$,$b$ and $c$ are shown in Fig.\ref{symmetryaxes}). Combining with inversion this gives $12$ elements divided into $6$ classes for $\text{D}_{3d}$ : ($E$), ($C_3,C_3^2$), ($C_{2,a}^{\prime},C_{2,b}^{\prime},C_{2,c}^{\prime}$), ($I$), ($I C_3,I C_3^2$) and ($I C_{2,a}^{\prime},I C_{2,b}^{\prime},I C_{2,c}^{\prime}$).

For the 3-sublattice dipolar and quadrupolar orders that we are interested in, we consider a single triangle as the unit cell.  The non-trivial transformations for the triangle formed by the points $1$,$2$,$3$ in Fig. \ref{symmetryaxes} which keep its centre (point x) fixed are $C_3$ and $\sigma_{h} C_{2}^{\prime}$ where $\sigma_{h}$ denotes reflection about the horizontal plane (the plane containing the triangular lattice). The axes $a^{\prime}$,$b^{\prime}$ and $c^{\prime}$ for the $C_2^{\prime}$ rotation are shown in Fig.\ref{symmetryaxes}. $\sigma_{h}$ is required with the $2$-fold rotation to bring the crystal field environment back to itself. Combining these with inversion about a lattice point generates the $12$ elements of $D_{3d}$ by up to translations by lattice vectors.

Denoting the transformations with an argument indicating the coordinate point that is left unchanged by the transformations ( eg. $C_3(1)$ indicates the operation of $C_3$ rotation keeping point $1$ fixed, $ I(1) C_3$(x) indicates the operation of $C_3$ rotation keeping point x fixed followed by inversion about the site now at location $1$), for the elements apart from identity and inversion, we have:
\begin{align*}
T_{-\mathbf{a}_1} C_3(\text{x})=&C_3(1)\\
T_{-\mathbf{a}_2} C_3^2(\text{x})=&C_3^2(1)\\
T_{-\mathbf{a}_1} \sigma_h C_{2,a^{\prime}}^{\prime}(\text{x})=&I(1) C_{2,a}^{\prime}(1)\\
\sigma_h C_{2,b^{\prime}}^{\prime}(\text{x})=&I(1) C_{2,b}^{\prime}(1)\\
T_{-\mathbf{a}_2} \sigma_h C_{2,c^{\prime}}^{\prime}(\text{x})=&I(1) C_{2,c}^{\prime}(1)\\
T_{\mathbf{a}_1} I(1) C_3(\text{x})=&I(1) C_3(1)\\
T_{\mathbf{a}_2} I(1) C_3^2(\text{x})=&I(1) C_3^2(1)\\
T_{\mathbf{a}_1} I(1) \sigma_h C_{2,a^{\prime}}^{\prime}(\text{x})=& C_{2,a}^{\prime}(1)\\
I(1)\sigma_h C_{2,b^{\prime}}^{\prime}(\text{x})=&C_{2,b}^{\prime}(1)\\
T_{\mathbf{a}_2} I(1) \sigma_h C_{2,c^{\prime}}^{\prime}(\text{x})=&C_{2,c}^{\prime}(1) ~.
\end{align*}
\subsection{Transformation of dipoles and quadrupoles under point group symmetries}
The symmetry transformations for irreps of the dipoles on the triangle of Fig. \ref{diagram_ugmodes} listed in Table  \ref{TableSpinsQuadrupoles}:
\begin{align*}
C_3:\quad m_{a}^{\alpha} \xrightarrow{}& m_{a}^{\alpha} \\ \nonumber
m_{e1}^{\alpha} \xrightarrow{}& -\frac{1}{2}m_{e1}^{\alpha} + \frac{\sqrt{3}}{2} m_{e2}^{\alpha} \\ \nonumber
m_{e2}^{\alpha} \xrightarrow{}& -\frac{\sqrt{3}}{2}m_{e1}^{\alpha} - \frac{1}{2} m_{e2}^{\alpha} \\ \nonumber
\sigma_h C_{2,a^{\prime}}^{\prime}:\quad m_{a}^{\alpha} \xrightarrow{}& m_{a}^{\alpha} \\ \nonumber
m_{e1}^{\alpha} \xrightarrow{}& -m_{e1}^{\alpha} \\ \nonumber
m_{e2}^{\alpha} \xrightarrow{}& m_{e2}^{\alpha} 
\label{symmetrym}
\end{align*} 
Further, inversion about a lattice point of a sublattice interchanges the other two sublattices. This action of interchange of two sublattices is the same as that of $\sigma_{h} C_{2}^{\prime}$ about the appropriate two fold axis. Under translation $T_{\mathbf{a}_1}$, sublattice $1\xrightarrow{}2\xrightarrow{}3\xrightarrow{}1$ which is also the action of $C_3$. Under translation $T_{\mathbf{a}_2}$, sublattice $1\xrightarrow{}3\xrightarrow{}2\xrightarrow{}1$ which is also the action of $C_3^2$. 
Under global SU(2) spin rotation, such that we have $m_{\widetilde{I}}^{\alpha} \xrightarrow{} R^{\alpha \beta} m_{\widetilde{I}}^{\beta}$ for $\widetilde{I}=a,e1,e2$, where $R^{\alpha \beta}$ are the SU(2) spin rotation matrices for spin-1. The dipoles are odd under time reversal, so $m_{\widetilde{I}}^{\alpha} \xrightarrow{} - m_{\widetilde{I}}^{\alpha}$. 

The symmetry transformations for irreps of the quadrupoles on the up triangle of Fig. \ref{diagram_ugmodes} listed in Table  \ref{TableSpinsQuadrupoles}:
\begin{eqnarray}
C_3:\quad Q_{a}^{\alpha\beta} &\xrightarrow{}& Q_{a}^{\alpha\beta} \\ \nonumber
Q_{e1}^{\alpha\beta} &\xrightarrow{}& -\frac{1}{2}Q_{e1}^{\alpha\beta} + \frac{\sqrt{3}}{2} Q_{e2}^{\alpha\beta} \\ \nonumber
Q_{e2}^{\alpha\beta} &\xrightarrow{}& -\frac{\sqrt{3}}{2}Q_{e1}^{\alpha\beta} - \frac{1}{2} Q_{e2}^{\alpha\beta} \\ \nonumber
\sigma_h C_{2,a^{\prime}}^{\prime}:\quad Q_{a}^{\alpha\beta} &\xrightarrow{}& Q_{a}^{\alpha\beta} \\ \nonumber
Q_{e1}^{\alpha\beta} &\xrightarrow{}& -Q_{e1}^{\alpha\beta} \\ \nonumber
Q_{e2}^{\alpha\beta} &\xrightarrow{}& Q_{e2}^{\alpha\beta}
\label{symmetryQ}
\end{eqnarray}
Like the case for dipolar modes, inversion about a lattice point of a sublattice interchanges the other two sublattices. This action of interchange of two sublattices is the same as that of $\sigma_{h} C_{2}^{\prime}$ about the appropriate two fold axis. Under translation $T_{\mathbf{a}_1}$, sublattice $1\xrightarrow{}2\xrightarrow{}3\xrightarrow{}1$ which is also the action of $C_3$. Under translation $T_{\mathbf{a}_2}$, sublattice $1\xrightarrow{}3\xrightarrow{}2\xrightarrow{}1$ which is also the action of $C_3^2$. Similarly, 
under spin rotations and time reversal, we respectively have $Q_{\widetilde{I}}^{\alpha\beta} \xrightarrow{} R^{\alpha \rho} R^{\beta \lambda} Q_{\widetilde{I}}^{\rho \lambda}$ and $Q_{\widetilde{I}}^{\alpha\beta} \xrightarrow{} Q_{\widetilde{I}}^{\alpha\beta}$ for $\widetilde{I}=a,e1,e2$.

\subsection{\label{Appendix:LandauNormalModes}Normal elastic modes and their transformations}
We consider the up triangle of Fig.\ref{diagram_ugmodes} with $\mathbf{u}_{1}$,$\mathbf{u}_{2}$ and $\mathbf{u}_{3}$ being the displacements from mean position. The three non-zero energy normal modes (in the basis \{ $u_{1x}$,$u_{1y}$,$u_{2x}$,$u_{2y}$,$u_{3x}$,$u_{3y}$\}) arranged as irreducible representations are given in Table \ref{TableNormalModes}.
\begin{table}
\begin{center}
 \begin{tabular}{| c | c |} 
 \hline
 Irrep & Eigenvector  \\ [0.5ex] 
 \hline
 $\epsilon_{a}$ & \{$\frac{-1}{2}$,$\frac{-1}{2 \sqrt{3}}$,$\frac{1}{2}$,$\frac{-1}{2 \sqrt{3}}$,$0$,$\frac{1}{\sqrt{3}}$\} \\ 
 \hline
 $\epsilon_{e1}$ & \{$\frac{-1}{2 \sqrt{3}}$,$\frac{-1}{2}$,$\frac{-1}{2 \sqrt{3}}$,$\frac{1}{2}$,$\frac{1}{\sqrt{3}}$,$0$\} \\
 $\epsilon_{e2}$ &  \{$\frac{-1}{2}$,$\frac{1}{2 \sqrt{3}}$,$\frac{1}{2}$,$\frac{1}{2 \sqrt{3}}$,$0$,$\frac{-1}{\sqrt{3}}$\} \\
\hline
\end{tabular}
\end{center}
\caption{Normal modes of an up triangle}
\label{TableNormalModes}
\end{table}
 Under the lattice symmetries, the normal modes transform as:
\begin{eqnarray}
C_3:\quad\epsilon_{a} &\xrightarrow{}& \epsilon_{a} \\ \nonumber
\epsilon_{e1} &\xrightarrow{}& -\frac{1}{2}\epsilon_{e1} + \frac{\sqrt{3}}{2} \epsilon_{e2} \\ \nonumber
\epsilon_{e2} &\xrightarrow{}& -\frac{\sqrt{3}}{2}\epsilon_{e1} - \frac{1}{2} \epsilon_{e2} \\ \nonumber
\sigma_h C_{2,a^{\prime}}^{\prime}:\quad \epsilon_{a} &\xrightarrow{}& \epsilon_{a} \\ \nonumber
\epsilon_{e1} &\xrightarrow{}& -\epsilon_{e1} \\ \nonumber
\epsilon_{e2} &\xrightarrow{}& \epsilon_{e2} 
\label{symmetryE}
\end{eqnarray}
The long wavelength strain modes, $\varepsilon_a, \varepsilon_{e1}, \varepsilon_{e2}$ (where $\varepsilon_a = \epsilon_a/a$, $\varepsilon_{e1} = \epsilon_{e1}/a$ and $\varepsilon_{e2} = \epsilon_{e2}/a$ ; $a=1$ is the lattice spacing), can be expressed in terms of Cartesian strains as given by Eq. \ref{strainsymmetry}. There are two independent elastic constants associated with the two modes as given in Eq. \ref{felastic}. Using the standard relations for the elastic constants for $D_{3d}$ symmetry \cite{articleLuthi} and comparing the harmonic elastic energy in written in terms of Cartesian strains with the Eq. \ref{felastic} we obtain,
\begin{align}
    c_a &= 2 (c_{11}+c_{12}) \nonumber\\
    c_e &= 2 (c_{11}-c_{12}) ~,
\end{align}
where the elastic constants on the right hand side are written in Voigt notation.
\bibliography{apssamp}

\end{document}